\documentclass[fleqn,usenatbib]{mnras}

\usepackage{newtxtext,newtxmath}

\usepackage[T1]{fontenc}

\DeclareRobustCommand{\VAN}[3]{#2}
\let\VANthebibliography\thebibliography
\def\thebibliography{\DeclareRobustCommand{\VAN}[3]{##3}\VANthebibliography}

\usepackage{graphicx}	
\usepackage{amsmath}

\usepackage{scalerel,tikz}
\usetikzlibrary{svg.path}
\definecolor{orcidlogocol}{HTML}{A6CE39}
\tikzset{orcidlogo/.pic={
 \fill[orcidlogocol] svg{M256,128c0,70.7-57.3,128-128,128C57.3,256,0,198.7,0,128C0,57.3,57.3,0,128,0C198.7,0,256,57.3,256,128z};
 \fill[white] svg{M86.3,186.2H70.9V79.1h15.4v48.4V186.2z}
 svg{M108.9,79.1h41.6c39.6,0,57,28.3,57,53.6c0,27.5-21.5,53.6-56.8,53.6h-41.8V79.1z M124.3,172.4h24.5c34.9,0,42.9-26.5,42.9-39.7c0-21.5-13.7-39.7-43.7-39.7h-23.7V172.4z}
 svg{M88.7,56.8c0,5.5-4.5,10.1-10.1,10.1c-5.6,0-10.1-4.6-10.1-10.1c0-5.6,4.5-10.1,10.1-10.1C84.2,46.7,88.7,51.3,88.7,56.8z};
}}
\newcommand\orcidicon[1]{\href{https://orcid.org/#1}{\mbox{\scalerel*{
\begin{tikzpicture}[yscale=-1,transform shape]
\pic{orcidlogo};
\end{tikzpicture}
}{|}}}}

\newcommand{\mach}{\mathcal{M}}
\newcommand{\msol}{\mathrm{M_\odot}}

\title[Impact of turbulent Mach number for the IMF]{The impact of the turbulent Mach number on star formation and the initial mass function}

\author[Mathew, Federrath, \& Seta]{
Sajay Sunny Mathew$^{\orcidicon{0000-0002-8381-8195}\,1}$\thanks{E-mail: \href{mailto:sajay.mathew@anu.edu.au}{sajay.mathew@anu.edu.au}},
Christoph Federrath$^{\orcidicon{0000-0002-0706-2306}\,1}$\thanks{E-mail: \href{mailto:christoph.federrath@anu.edu.au}{christoph.federrath@anu.edu.au}}, and
Amit Seta$^{\orcidicon{0000-0001-9708-0286}\,1}$\thanks{E-mail: \href{mailto:amit.seta@anu.edu.au}{amit.seta@anu.edu.au}}
\\
$^{1}$Research School of Astronomy and Astrophysics, Australian National University, Canberra, ACT~2611, Australia
}
\date{Accepted 23 September 2026. Received 21 August 2026; in original form 22 October 2025}

\pubyear{\the\year{}}

\begin{document}
\label{firstpage}
\pagerange{\pageref{firstpage}--\pageref{lastpage}}
\maketitle

\begin{abstract}
Turbulence regulates star formation by influencing the density structure and fragmentation of molecular clouds, and therefore it is expected to play a key role in setting the initial mass function (IMF). We study how the strength of the turbulent shocks affects star formation and the IMF by comparing a series of magnetohydrodynamical (MHD) simulations of star cluster formation in clouds with three different rms Mach number values, $\mathcal{M}=2.5, 5$, and $10$, but otherwise the same Alfv\'en Mach number ($\mathcal{M}_\mathrm{A}\sim3$) and virial parameter ($\alpha_\mathrm{vir}\sim0.5$). All three simulations include stellar feedback in the form of protostellar jets/outflows and accretion heating. We find that the star formation rate per freefall time ($\mathrm{SFR_{ff}}$) for the $\mathcal{M}=10$ model is higher by a factor of $\sim3$ compared to the $\mathcal{M}=2.5$ and $5$ cases, which have similar $\mathrm{SFR_{ff}}$. In terms of the IMF, the $\mathcal{M}=5$ and $10$ models produce almost similar distributions, resembling typical observed IMFs. The $\mathcal{M}=2.5$ model, on the other hand, produces a bimodal IMF with a primary peak at supersolar masses ($\sim 2\, \mathrm{M_\odot}$) and a secondary peak in the substellar regime ($< 0.1\, \mathrm{M_\odot}$). The $\mathcal{M}=2.5$ and $5$ simulations producing significantly different IMFs while having similar $\mathrm{SFR_{ff}}$, suggest that a distinct set of physical processes governs the SFR and IMF. Through a resolution study for the $\mathcal{M}=2.5$ case, we find our base models have reached near numerical convergence, potentially only slightly underestimating close-binary formation. We also see signatures of bimodality in the $\mathcal{M}=2.5$ model for the multiplicity fraction and stellar angular momentum distribution.   
\end{abstract}




\section{Introduction}
Supersonic turbulence within the interstellar medium (ISM) plays a pivotal role in the star formation process \citep[e.g.,][]{1993ApJ...419L..29E,1995MNRAS.277..377P,2000ApJ...535..887K,2004ARA&A..42..211E,sfr_fk2012,2012ApJ...754...71K,2013ApJ...770..150H,2022MNRAS.514..957S}. The low star formation rate observed in molecular clouds has been primarily attributed to the complex interplay between gravity, turbulence, magnetic fields, and stellar feedback \citep{sfr_km2005,2006ApJ...640L.187L,vazquezsemadeni2007,2009ApJS..181..321E,2010ApJ...709...27W,2011MNRAS.411...65B,2011ApJ...729..133M,2013MNRAS.435.1701C,2014ApJ...790..128F,2014prpl.conf..451F,2015MNRAS.450.4035F,2016ApJ...833..229L,2016MNRAS.458.1671K,2016ApJ...831...73V,2017ApJ...847..104O,2018MNRAS.473.4975T,2019A&A...622A.125L,2019FrASS...6....7K,mathew2021,2021MNRAS.502.3646G,2021MNRAS.506.3239G,2021NatAs...5..365F,2024A&A...683A..13L,2024A&A...690A..44B,2025MNRAS.tmp.1946V}. In addition to suppressing star formation on global scales by moving the gas around \citep{2024JApA...45...17S}, turbulence also induces local compressions, producing high-density regions that can potentially undergo collapse to form stars \citep{2001ApJ...553..227P,2004RvMP...76..125M,2008ApJ...684..395H,sfr_fk2012}. The degree of gas compression, which influences the fragmentation and gravitational collapse, depends on the strength of the turbulent shocks within the clouds. Therefore, the Mach number of turbulence, which defines this strength and is given by the ratio of the turbulent gas velocity dispersion, $\sigma_\mathrm{v}$, to the sound speed, $c_{\mathrm{s}}$, is a key cloud parameter that controls star formation \citep{1994ApJ...423..681V,1997MNRAS.288..145P,2007ApJ...665..416K,sfr_fk2012,hopkins2013}.

An important feature of star formation is the similarity observed in the mass distribution of new star clusters across a wide range of environments \citep{2010ARA&A..48..339B,2014prpl.conf...53O,2014MNRAS.444.1957D,2017MNRAS.464.1738D,2018PASA...35...39H,2020SSRv..216...70L}. This distribution is referred to as the initial mass function (IMF). The IMF follows a power law at the high-mass end and can be represented as $dN \propto M^{-\Gamma}\, d\mathrm{log}M\, (M > 1\, \mathrm{M_\odot})$, where $N$ is the number of stars, $M$ is the stellar mass, and $\Gamma\sim1.35$ \citep{1955ApJ...121..161S}. At subsolar masses, the IMF can be described by a three-segment power-law \citep{2001MNRAS.322..231K} or a log-normal function \citep{chabrier2005}. The peak mass or the characteristic mass of the IMF lies around $0.2$--$0.3\, \mathrm{M_\odot}$ \citep{2003PASP..115..763C,2008ApJ...681..365E,2014prpl.conf...53O}. However, there is a non-negligible scatter in the peak mass and the power-law exponent at the high-mass end, as obtained in different observational studies \citep{2014MNRAS.444.1957D,2023Natur.613..460L,2023ApJ...959...88D,2024ARA&A..62...63H}. 

The role of turbulence in determining the IMF is yet to be fully understood because of the highly complex and non-linear nature of turbulent flows. Previous numerical works looked at the effect of the turbulent Mach number by varying the velocity, but not necessarily keeping other important parameters fixed that also depend on the velocity dispersion. Such studies find that $\mathcal{M}$ either does not significantly affect the IMF shape or causes only minor variations \citep{sfr_km2005,2006ApJ...637..384B,2015MNRAS.452..566B,2018A&A...611A..88L}. However, varying the velocity dispersion also changes the virial parameter $\alpha_{\mathrm{vir}}$, i.e., the ratio of twice turbulent kinetic to gravitational energy. Variations in $\alpha_{\mathrm{vir}}$ modify the cloud evolution, the fragmentation process, and the IMF \citep{2008MNRAS.386....3C,sfr_fk2012,padoan2012,Haugb_lle_2018,2021ApJ...911..128K,mathew2025}. Hence, the results in these studies likely reflect the combined effect of changing both $\mathcal{M}$ and $\alpha_{\mathrm{vir}}$. To understand the overall influence of turbulence, it is therefore crucial to isolate the effects of different turbulence properties, not only $\mathcal{M}$ and $\alpha_{\mathrm{vir}}$, but also the turbulence driving parameter $b$, which is another important quantity that significantly impacts the IMF \citep{2010A&A...516A..25S,2015MNRAS.449..662L,2017MNRAS.465..105L,mathew2023}.

In \citet{mathew2023}, we performed a detailed study on the influence of the turbulence driving ($b$) on the IMF, keeping all other turbulent properties fixed. Similarly, in \citet{mathew2025}, we investigated the individual effect of $\alpha_{\mathrm{vir}}$. In this work, keeping $b$ and $\alpha_{\mathrm{vir}}$ fixed, we analyse the role of the large-scale turbulent Mach number ($\mathcal{M}$) of the cloud. We perform a series of numerical simulations of star cluster formation with $\mathcal{M}=2.5, 5$, and $10$, and study how the cloud morphology, SFR, IMF, multiplicity, and angular momentum vary between them. The simulations incorporate crucial physical mechanisms, including gravity, turbulence, magnetic fields, and protostellar feedback via radiative heating and mechanical outflows \citep[see][for further details]{mathew2021,mathew2023,mathew2024}.

The paper is structured as follows. In Section~\ref{sec:methods}, we provide details of the simulation setup and the initial conditions. In Section~\ref{sec:results}, we study how the stellar properties, such as the SFR, IMF, binary characteristics, and angular momentum, vary between our simulation models with different cloud Mach numbers. The main conclusions are summarised in Section~\ref{sec:conclusions}.

\section{Methodology}
\label{sec:methods}
\subsection{Numerical setup}
We use the simulation framework in \citet{mathew2021} and \citet{mathew2023} for this study. Here, we give a short overview of the numerical setup employed. Detailed specifications of the simulation configuration are outlined in \S2 of \citet{mathew2021}. To model star cluster formation through cloud collapse, we solve the magnetohydrodynamical (MHD) equations in the presence of gravity using a highly modified version of the \textsc{flash} (version~4) code \citep{2000ApJS..131..273F,2008ASPC..385..145D} with the positivity-preserving HLL3R Riemann solver \citep{2011JCoPh.230.3331W}. The gas in the cloud is magnetised (see initial conditions below) and turbulence is continuously driven \citep[see][and the references therein]{2010A&A...512A..81F,2022ascl.soft04001F,mathew2021,2021NatAs...5..365F}.

We utilise sink particles to model the densest and gravitationally bound regions of the gas, which is a standard procedure in numerical star formation studies \citep{1995MNRAS.277..362B,2004ApJ...611..399K,2010ApJ...713..269F}. We use the sink particle prescription developed in \citet{2010ApJ...713..269F}. In this implementation, when a spherical volume of gas with radius $r_\mathrm{sink}$ satisfies a range of sink formation criteria, the enclosed gas above the sink density threshold ($3.81\times10^{-16}\, \mathrm{g\,cm^{-3}}$) will be replaced by a new sink particle or accreted onto existing sink particles within $r_\mathrm{sink}$. We adopt a sink particle radius of $r_\mathrm{sink} = 250\, \mathrm{AU}$. Here, explicit checks are carried out to ensure that the accreted gas is gravitationally bound and undergoing collapse \citep{2010ApJ...713..269F}. We note that, while the sink particles gravitationally interact with each other, they are not allowed to merge. We use sink particles to represent individual stars. While stellar mergers are possible in high-mass star formation under extreme conditions \citep{2007ARA&A..45..481Z}, they are not expected in our simulations, which represent Milky Way-like, low-mass star-forming environments \citep{2010ApJ...713..269F,2025arXiv251012203F}. To model stellar radiative heating and jets/outflows, we use sub-grid models developed in \citet{mathew2020} and \citet{2014ApJ...790..128F}, respectively. Incorporating these feedback mechanisms into star cluster formation simulations is essential to accurately capture the fragmentation process and reproduce the IMF \citep{2006ApJ...640L.187L,bate2009,Li_2010,2011ApJ...740...74K,2011ApJ...740...36N,2012ApJ...754...71K,2014ApJ...790..128F,2016MNRAS.458..673G,2017JPhCS.837a2007F,2018MNRAS.476..771C,mathew2020,hennebelle2020,2021MNRAS.502.3646G,2021MNRAS.500.3594R,mathew2021}. 
 
\subsection{Initial conditions}
\label{sec:initial conditions}

\begin{table*}
\caption{Initial conditions and key measurements of each simulation model.}
\renewcommand{\arraystretch}{1.0}
\setlength{\tabcolsep}{3.0pt}
\label{tab:init}
\begin{tabular}{ccccccccccccccc}
\hline
\hline
 Model & $L\, [\mathrm{pc}]$ & $\rho_0\, [\mathrm{g\, cm^{-3}}]$ & $M_{\mathrm{cl}}\, [\mathrm{M_\odot}]$ & $M_{\mathrm{cl}}/M_{\mathrm{J}}$ & $\alpha_{\mathrm{vir}}$ & $\mathcal{M}$ & $B_0\, [\mathrm{\mu G}]$ & $\mathcal{M_{\mathrm{A}}}$ & $\mu_{\mathrm{B}}$ & $\beta$ & $N_{\mathrm{sims}}$ & $N_{\mathrm{tot}}$ & $M_{\mathrm{avg}}\, [\msol]$ & $j_{\mathrm{mean}}\, [\mathrm{cm^2\, s^{-1}}]$\\
 (1) & (2) & (3) & (4) & (5) & (6) & (7) & (8) & (9) & (10) & (11) & (12) & (13) & (14) & (15)\\
\hline
$\mathcal{M} = 2.5$ & 2 & $1.64\times10^{-21}$ & 194 & 11.5 & 0.5 & 2.5 & 2.5 & 2.9 & 6.9 & 2.7 & 29 & 206 & 1.4 & $1.5\times10^{19}$\\
$\mathcal{M} = 5$ & 2 & $6.56\times10^{-21}$ & 775 & 92.4 & 0.5 & 5 & 10 & 2.9 & 6.9 & 0.7 & 5 & 223 & 0.9 & $1.2\times10^{19}$\\
$\mathcal{M} = 10$ & 2 & $2.62\times10^{-20}$ & 3097 & 737.7 & 0.5 & 10 & 40 & 2.9 & 6.9 & 0.2 & 1 & 231 & 0.7 & $8.8\times10^{18}$\\
\hline
$\mathcal{M} = 2.5$, $\alpha_{\mathrm{vir}} = 0.125$ & 2 & $6.56\times10^{-21}$ & 775 & 92.4 & 0.125 & 2.5 & 5 & 2.9 & 13.8 & 2.7 & 5 & 236 & 0.8 & $1.0\times10^{19}$\\
\hline
\hline
\end{tabular}
\\
\raggedright\textbf{Notes.} Column~1: simulation model name. Column~2: size of cloud/computational domain. Column~3: initial mean gas density. Column~4: total mass. Column~5: number of Jeans masses. Column~6: virial parameter. Column~7: rms sonic Mach number. Column~8: initial magnetic field strength. Column~9: Alfv\'en Mach number. Column~10: mass-to-flux ratio. Column~11: plasma beta (thermal-to-magnetic pressure ratio). Column~12: number of simulations (number of random seeds for the turbulence driving) performed. Column~13: total number of star particles formed. Column~14: average stellar mass. Column~15: mean specific angular momentum of stars.
\end{table*}

Each simulation is performed in a three-dimensional numerical domain with side length $L = 2\, \mathrm{pc}$, having periodic boundary conditions and an adaptive mesh refinement (AMR) framework for the computational grids. The maximum effective resolution of $4096^3$ cells or equivalently a minimum cell size of $100\,$AU is employed in all simulations, except for the additional simulations carried out as part of the resolution study in \S\ref{sec:res_study}. A uniform gas density $\rho_0$ and a magnetic field strength $B_0$ directed only along the $z$-axis are initially set in the simulations. The cloud is initially isothermal at 10\,K, but a polytropic equation of state (EOS) is used, making the gas pressure, and consequently the temperature via the ideal gas law, dependent on the local density \citep[see][for further information about the polytropic EOS used here]{mathew2021}. The polytropic EOS is based on previous radiation-hydrodynamical simulations and theoretical works \citep{1969MNRAS.145..271L,1993ApJ...411..274Y,2000ApJ...531..350M,2009ApJ...703..131O}. We adopt a turbulence driving parameter value of $b \sim 0.4$. The turbulence driving parameter $b$ is a measure of the ratio of energies in compressive and solenoidal driving modes \citep[see][for a detailed explanation of the driving parameter]{2008ApJ...688L..79F}. $b \sim 0.4$ aligns with a natural mixture of turbulence driving modes, i.e., equal power in compressive and solenoidal driving modes \citep[see][for details on the turbulence driving scheme]{2008ApJ...688L..79F,2010A&A...512A..81F,2022ascl.soft04001F}. The turbulence driving creates high-density regions in the form of filaments and clumps, developing via turbulent shocks. The turbulent gas flow makes the magnetic field structure inhomogeneous due to the compression, twisting, and stretching of magnetic field lines \citep{1998MNRAS.294..718S,2005PhR...417....1B,2016JPlPh..82f5301F,2020PhRvF...5d3702S,2021PhRvF...6j3701S}, generating a field structure similar to that observed in molecular clouds \citep{1992A&A...257..715F,2008A&A...487..247F,2017ApJ...847...92H,2016A&A...586A.138P}.

Turbulence is driven in the simulations without self-gravity for two turbulent crossing times ($2\,t_\mathrm{turb}=L/(\mathcal{M}c_\mathrm{s}) = 2\,\mathrm{Myr}$ ) to ensure that it fully develops and reaches a steady state \citep{2010A&A...512A..81F,pricefederrath2010}. When a steady state of turbulence is reached, we turn on self-gravity, which we define as time $t = 0$. The simulations are allowed to evolve until a star formation efficiency (SFE) of 5\% is reached, i.e., when 5\% of the total cloud mass has been accreted into stars. This choice is motivated by the expectation that, at roughly this SFE, radiative feedback processes (stellar winds and supernovae), which are not included in the present simulation setup, become effective at dispersing the cloud and thereby halting further star formation \citep{2020MNRAS.493.2872C,2022MNRAS.512..216G}.

To study the influence of the large-scale turbulent Mach number on the star formation process, we carry out simulations with $\mathcal{M} = 2.5, 5$ and $10$ by varying $\sigma_{\mathrm{v}}$, which is the velocity dispersion on the driving scale. The sound speed is set to $c_{\mathrm{s}} = 0.2\, \mathrm{km\, s^{-1}}$. Changing the velocity dispersion to vary $\mathcal{M}$ would also affect $\alpha_{\mathrm{vir}}$ and the Alfv\'en Mach number $\mathcal{M_\mathrm{A}}$. However, for a controlled parameter study that only varies $\mathcal{M}$, we adjust the density and magnetic field strength to ensure that $\alpha_{\mathrm{vir}}$ and $\mathcal{M_\mathrm{A}}$, as well as the mass-to-flux ratio $\mu_{\mathrm{B}}$, remain the same. The $\alpha_{\mathrm{vir}}$, $\mathcal{M_\mathrm{A}}$, and $\mu_{\mathrm{B}}$ in all three simulation models are fixed at $0.5$, $2.9$, and $6.9$, respectively. Keeping the above quantities constant distinguishes our work from previous numerical studies on the role of the turbulent Mach number on the IMF. As the governing physical processes are determined by these relative, dimensionless quantities, it is crucial to keep them fixed when varying $\mathcal{M}$. Nevertheless, it should be noted that changing $\mathcal{M}$ without affecting any of the other cloud parameters is not feasible, as many of them are mathematically and physically linked. For example, the number of Jeans masses $M_{\mathrm{cl}}/M_{\mathrm{J}}$ changes on varying $\mathcal{M}$ at fixed $\alpha_{\mathrm{vir}}$ (see Appendix~\ref{sec: isolate_proof}). We also carried out additional simulations where we lower $\mathcal{M}$ and keep $M_{\mathrm{cl}}/M_{\mathrm{J}}=92.4$ as in the main $\mathcal{M}=5$ model, which required increasing $\mu_{\mathrm{B}}$, and lowering $\alpha_{\mathrm{vir}}$, again due to the interdependence of these parameters (Appendix~\ref{sec: isolate_proof}). The initial values of all relevant physical parameters in the three main simulation models and the additional simulation model are listed in Tab.~\ref{tab:init}.

\section{Results}
\label{sec:results}
We carried out a series of cloud-collapse simulations for three different values of the large-scale turbulent Mach number: $\mathcal{M} = 2.5, 5$, and $10$. For $\mathcal{M} = 2.5$, we performed $29$~simulations with different turbulence realisations, yielding a total of $206$~star particles. For $\mathcal{M} = 5$ and $10$, we have $5$ and $1$~simulation(s) each, producing $223$ and $231$~star particles, respectively. We note that the data for $\mathcal{M} =5$ is taken from the simulations in \citet{mathew2021}. The choice of the number of simulations in each model is based on the aim of forming a similar number of stars in all three models, for a reasonable comparison based on similar statistics.

\begin{figure*}
    \centering
    \includegraphics[width=\textwidth]{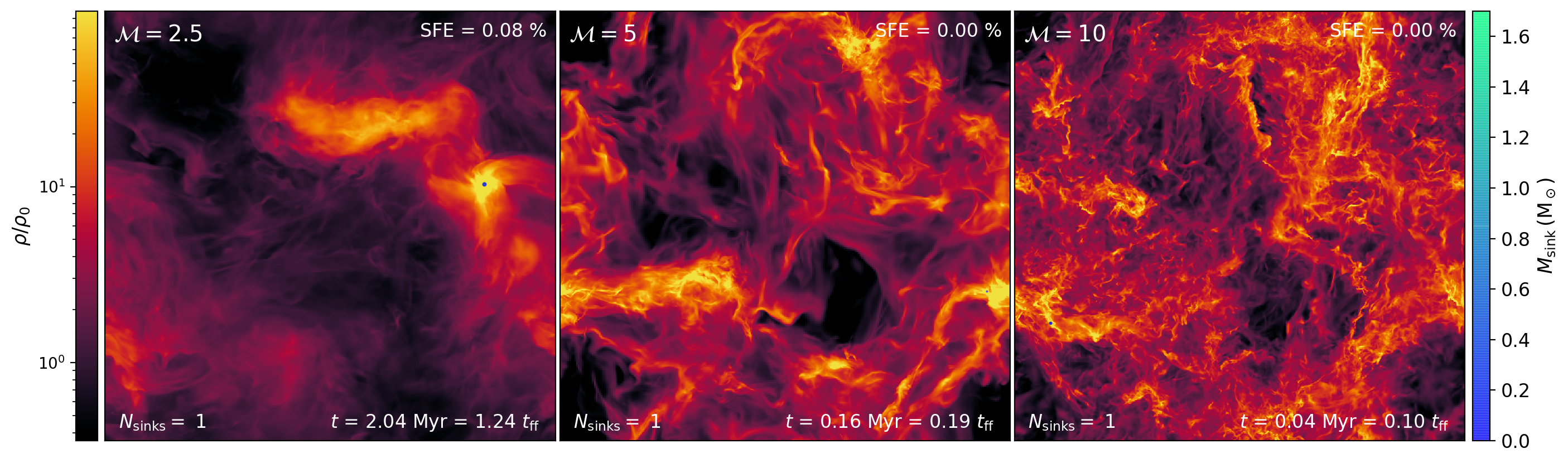}
    \caption{Mass-weighted gas density projection divided by the corresponding initial density for $\mathcal{M} = 2.5$ (left panel), $5$ (middle panel), and $10$ (right panel), at the moment the first sink particle (star+disc system) forms ($\mathrm{SFE}\approx0$). The circular markers correspond to the sink particle positions, and the colour bar on the right shows the mass of the sink particles. The size of the markers is scaled to the sink particle masses.} 
    \label{fig:densmap1}
\end{figure*}

Fig.~\ref{fig:densmap1} shows the density maps for the three models around the time the first star forms. It is evident that there is a higher prevalence of high-density, large-scale structures for the high-$\mach$ cases, which have the potential to form stars, because of the increased strength of the turbulent shocks \citep{2006ApJ...637..384B,sfr_fk2012,2026A&A...706A.178N}. Consequently, the density contrast within these structures increases with increasing $\mach$ \citep{2005ApJ...630L..45K,2012MNRAS.423.2680M,2021MNRAS.507.4335K}. The turbulent kinetic energy in the $\mach = 10$ case is high enough to generate strong shocks even on the smallest scales, causing the highly feathered morphology seen in this case.

\begin{figure*}
    \centering
    \includegraphics[width=\textwidth]{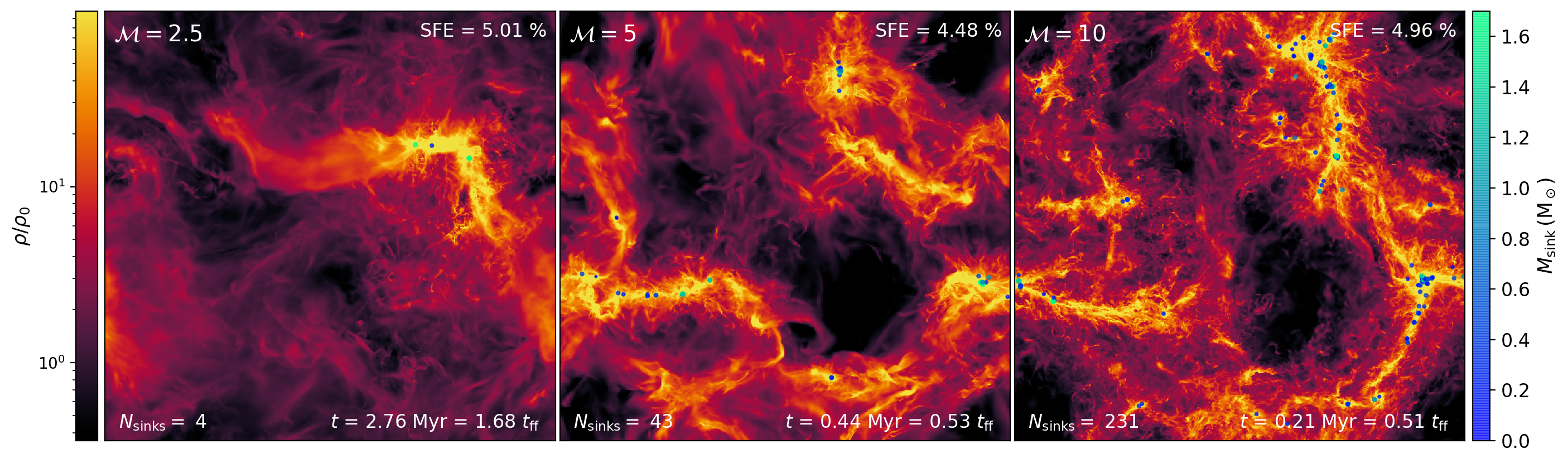}
    \caption{Same as Fig.~\ref{fig:densmap1}, but at star formation efficiency (SFE) of $\approx 5$\%. Note that some of the multiple systems will appear as nearly concentric markers (circles with central dots), as the separation between their members are too small to be visually distinguished in the figure.}
    \label{fig:densmap2}
\end{figure*}

Fig.~\ref{fig:densmap2} shows corresponding density maps at SFE = 5\%. The number of stars formed increases with increasing $\mach=2$, $5$, and $10$ producing $4$, $43$, and $231$ stars, respectively. Thus, more turbulent fragmentation is a consequence of the higher density contrasts induced with increasing $\mach$. 

\subsection{Star formation rate}
\label{sec:sfr}
\begin{figure}
    \includegraphics[width=\columnwidth]{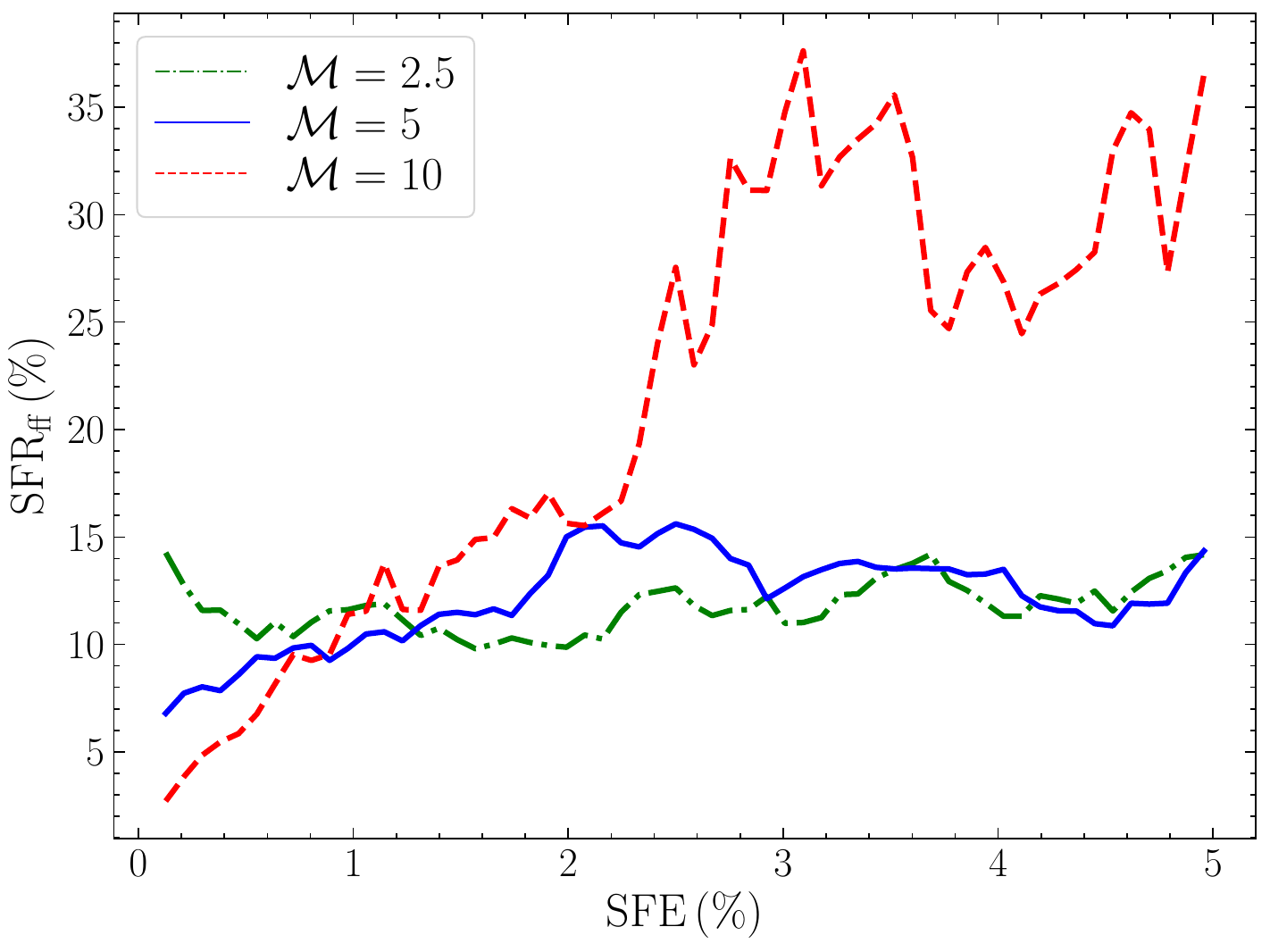}
    \caption{Star formation rate per freefall time ($\mathrm{SFR_{ff}}$) as a function of SFE in simulations with $\mathcal{M} = 2.5$ (dash-dotted), $\mathcal{M} = 5$ (solid), and $\mathcal{M} = 10$ (dashed).}  
    \label{fig:sfr}
\end{figure}

The evolution of the star formation rate per freefall time ($\mathrm{SFR_{ff}}$) for the three simulation models is shown in Fig.~\ref{fig:sfr}. Here, $\mathrm{SFR_{ff}} = (\mathrm{SFE(\%)}\,/\,t)\times\,t_{\mathrm{ff}}(\rho_0)$, where $t_{\mathrm{ff}}(\rho_0)$ is the freefall time at the mean gas density $\rho_0$ in the corresponding simulation. $\mathrm{SFR_{ff}}$ for $\mathcal{M} = 10$ is significantly higher compared to that of $\mathcal{M} = 2.5$ and $\mathcal{M} = 5$. This is qualitatively consistent with turbulence-regulated theories of star formation \citep{sfr_km2005,sfr_pn2011,sfr_hc2011,sfr_fk2012}. However, it is to be noted that a few of the dimensionless quantities vary between the different Mach number runs because of their interdependence with $\mathcal{M}$ (cf.~Appendix~\ref{sec: isolate_proof}), whereas they are assumed to be fixed in the theoretical predictions.

$\mathrm{SFR_{ff}}$ in the $\mathcal{M} = 2.5$ and $5$ models reaches a quasi-steady state around SFE = 1\%, whereas $\mathrm{SFR_{ff}}$ in the $\mathcal{M} = 10$ model continues to grow and only begins to reach a quasi-steady state at around SFE = 3\%. The $\mathrm{SFR_{ff}}$ for $\mathcal{M} = 2.5$ and $5$ is similar with an average value of ($13\pm1)\%$ in the \mbox{$\mathrm{SFE}=3$--$5\%$} range. The average $\mathrm{SFR_{ff}}$ for $\mathcal{M} = 10$ in the same SFE range is ($31\pm4)\%$. The average stellar accretion rates for $\mathcal{M}=2.5$, $5$, and $10$ are $4.1\times10^{-6}$, $4.9\times10^{-6}$, and $1.3\times10^{-5}$ $\mathrm{M_\odot}\, \mathrm{yr^{-1}}$, respectively. The higher $\mathrm{SFR_{ff}}$ in the $\mathcal{M}=10$ model is therefore not only due to the higher number of stars forming (higher degree of fragmentation) but also due to the higher accretion rate of the newly formed stars, as a consequence of the higher dense gas fraction when $\mach$ is increased \citep{sfr_fk2012}.

The time at which the first star particle forms varies between the three models, with the first star particle forming in the simulations with $\mathcal{M} = 2.5, 5$, and $10$ at $1.68\, \mathrm{Myr} = 1.02\, t_{\mathrm{ff}}$, $0.41\, \mathrm{Myr} = 0.50\, t_{\mathrm{ff}}$, and $0.03\, \mathrm{Myr} = 0.09\, t_{\mathrm{ff}}$, respectively, averaged over all turbulence realisations of the respective model. Since the clouds with $\mathcal{M} = 2.5$ have undergone significant time evolution before the first star forms, gravity had sufficient time to accumulate gas around the over-dense regions. At the onset of star formation, all this accumulated gas is rapidly accreted by the forming star(s), leading to high accretion rates initially. This explains why $\mathrm{SFR_{ff}}$ is high for $\mathcal{M} = 2.5$ at $\mathrm{SFE}\sim0$. 

\subsection{Time of formation}
\label{sec:formtime}
\begin{figure}
    \includegraphics[width=\columnwidth]{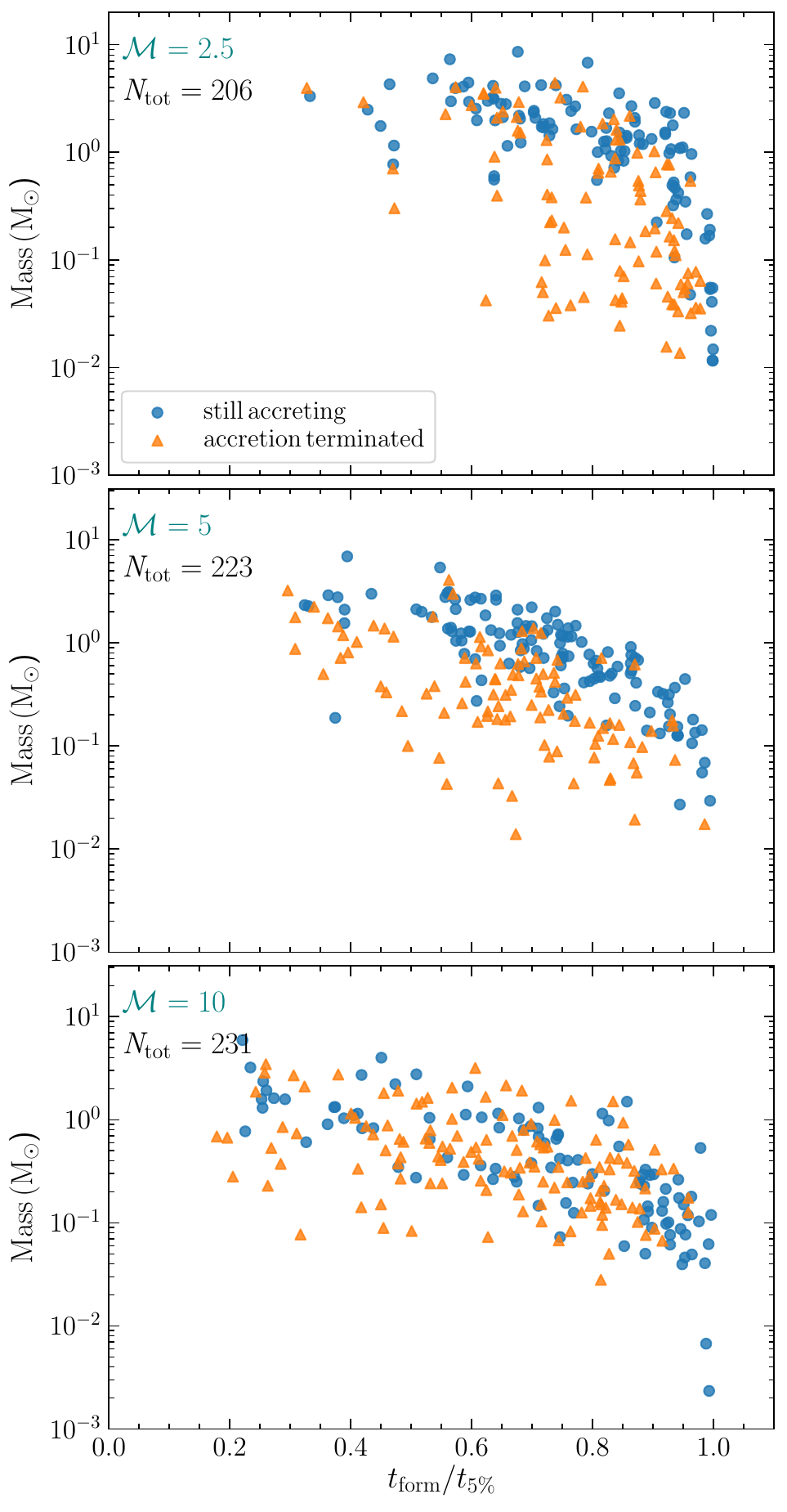}
    \caption{Mass of stars and brown dwarfs as a function of their time of formation, with $t_{\mathrm{5\%}}$ representing the time corresponding to $\mathrm{SFE} = 5\%$, for $\mach=2.5$, $5$, and $10$ (top to bottom panels). The circular markers represent the objects that are still accreting at the end of the simulation, while the triangular markers correspond to the stars that have stopped accreting.}
    \label{fig:tformvsmass}
\end{figure}

Fig.~\ref{fig:tformvsmass} shows the relationship between the final mass of star particles (mass at $\mathrm{SFE} = 5\%$) and their time of formation, with $t_{\mathrm{form}}/t_{5\%}$ representing the time of formation in units of the time taken by the simulation to reach $\mathrm{SFE} = 5\%$. We find a weak negative correlation between the mass of a star and the time it forms. In all three models, most of the high-mass stars form in the early stages of the cluster formation process, which has also been observed in previous simulations \citep[e.g.,][]{bate2009,2012MNRAS.419.3115B}. Further, a large fraction of the very-low-mass population ($M < 0.1\, \mathrm{M_\odot}$) forms towards the late stages of the simulation. 

We see that, while stars start to form soon after gravity is turned on ($t=0$) in the case of $\mathcal{M}=5$ and $10$, the first stars form much later for $\mathcal{M}=2.5$. Further, in the $\mathcal{M}=5$ and $10$ simulations, the stars that form at $t_{\mathrm{form}}/t_{5\%}\sim0.6-0.8$ generally have a final mass in the range $0.1 < M\, (\mathrm{M_\odot}) < 1$. On the other hand, the star particles that form around the same time in the $\mathcal{M}=2.5$ simulations only reach substellar masses ($M < 0.1\, \mathrm{M_\odot}$). These star particles have also stopped accreting (shown with triangular markers). The above differences in the dynamical evolution between the low-Mach and high-Mach cases further emphasise the role of the strength of turbulence and also contribute to variations in the IMF, which is discussed next.     

\subsection{Initial mass function}
\label{sec:imf}
\begin{figure*}
    \centering
    \includegraphics[width=1.8\columnwidth]{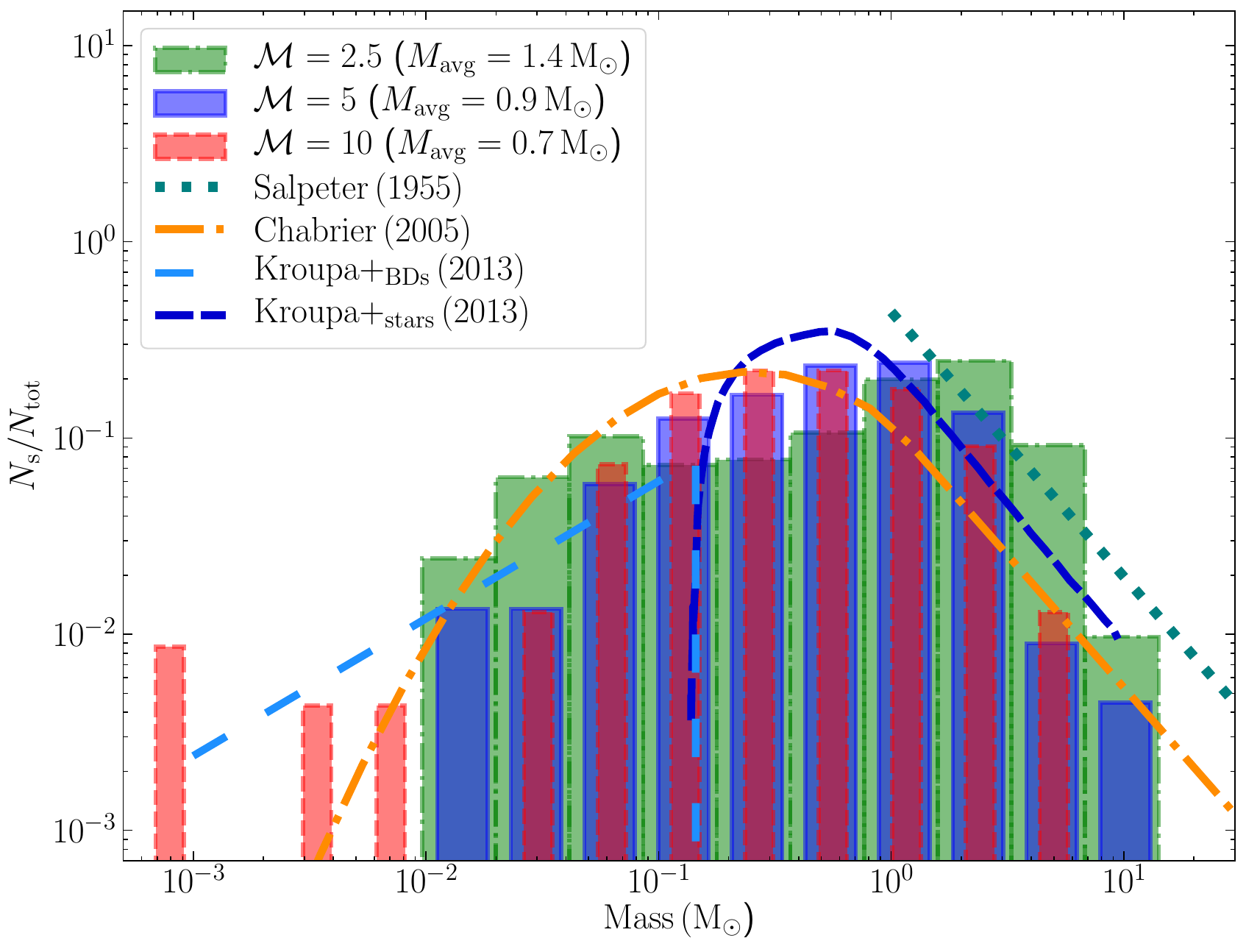}
    \caption{Stellar mass function for simulations with $\mathcal{M} = 2.5$ (green histogram with dash-dotted edges), $5$ (blue histogram with solid edges), and $10$ (red histogram with dashed edges) at SFE = 5\%. Each bin represents the ratio of the number of star particles in the associated mass range ($N_{\mathrm{s}}$) to the total number of star particles ($N_{\mathrm{tot}}$). The plotted curves represent system IMF models from observations, by \citet{1955ApJ...121..161S} (dotted), \citet{chabrier2005} (dash-dotted), \citet{2013pss5.book..115K} for brown dwarfs (long-dashed) and stars (short-dashed).}  
    \label{fig:imf}
\end{figure*}

Fig.~\ref{fig:imf} shows the mass distribution of stars in the models with $\mathcal{M} = 2.5$ (green histogram with dash-dotted edges), $5$ (blue histogram with solid edges), and $10$ (red histogram with dashed edges). The dotted and dash-dotted lines correspond to the \citet{1955ApJ...121..161S} and \citet{chabrier2005} IMFs, respectively. The long-dashed and short-dashed lines represent the \citet{2013pss5.book..115K} system IMFs for brown dwarfs and stars, respectively. The shape of the distributions for $\mathcal{M}=5$ and $10$ agrees with the general shape of \citet{chabrier2005} and \citet{2013pss5.book..115K} IMFs, although their peaks are at slightly different masses. The average stellar mass in the $\mathcal{M}=5$ and $10$ models are $0.9\, \msol$ and $0.7\, \msol$, respectively. The mass distribution for $\mathcal{M} = 2.5$ is significantly different from the observational IMF models \citep{chabrier2005,2013pss5.book..115K}, as well as from those for $\mathcal{M}=5$ and $10$. The average stellar mass in the $\mathcal{M}=2.5$ model ($1.4\, \msol$) is higher by a factor of $\sim2$ than in the $\mathcal{M}=10$ model. The increase in the mean stellar mass with decreasing $\mach$ is consistent with the current theories of the IMF based on gravo-turbulent fragmentation \citep{2002ApJ...576..870P, 2008ApJ...684..395H,2009ApJ...702.1428H,2012MNRAS.423.2037H,2013MNRAS.430.1653H}. However, it is important to note that the form of the IMF for $\mathcal{M}=5$ and $10$ is relatively similar and a noticeable difference emerges only when the Mach number is reduced to $\mathcal{M}=2.5$. This indicates that the IMF has a non-linear dependence on the Mach number. Interestingly, while the $\mathcal{M}=2.5$ and $5$ models exhibit similar trends in $\mathrm{SFR_{ff}}$ (c.f., Fig.~\ref{fig:sfr}), it is the $\mathcal{M}=5$ and $10$ models that are more similar in terms of the IMF. This shows that the SFR and IMF respond differently to changes in the turbulent Mach number. 

\begin{table}
	\caption{Fraction of stars in different mass ranges.}
    \renewcommand{\arraystretch}{1.5}
	\label{tab:m_frac}
	\begin{tabular}{c|ccc|} 
	    \hline
		\hline
		 Model & \multicolumn{3}{c|}{Fraction of stars in mass range}\\
         \cline{2-4}
         & $M \leq 0.1\, \mathrm{M_\odot}$ & $0.1 < M\, (\mathrm{M_\odot}) < 1$ & $M \geq 1\, \mathrm{M_\odot}$\\
		\hline
		\hline 
		$\mathcal{M}=2.5$  & $0.20$ & $0.30$ & $0.50$\\
		$\mathcal{M}=5$ & $0.09$ & $0.59$ & $0.32$ \\
	    $\mathcal{M}=10$ & $0.13$ & $0.66$ & $0.21$ \\
		\hline
	\end{tabular}
\end{table}

The IMF in the $\mathcal{M}=2.5$ model cannot be reproduced by a simple shift in the peak of the IMF in the higher Mach number cases or by a change in their width. Instead, it exhibits a fundamentally different form. The simulations with $\mathcal{M}=2.5$ have a higher fraction of supersolar masses compared to the other two models, but a significantly lower fraction of stars in the mass range $0.1 < M\, (\mathrm{M_\odot}) < 1$. The fraction of stars in different mass ranges for the three models is listed in Tab.~\ref{tab:m_frac}. An important distinction in the $\mathcal{M}=2.5$ case is that the shape of the IMF is bimodal, i.e., there exists a primary peak at $\sim2\, \mathrm{M_\odot}$ and a secondary peak around $0.06-0.07\, \mathrm{M_\odot}$. A similar bimodal form of the IMF has been found in observations \citep{2013pss5.book..115K,Drass_2016}, which may be related to turbulent properties of these clouds, possibly a relatively low turbulent Mach number, as suggested by the present work. 

\begin{figure}
    \includegraphics[width=\columnwidth]{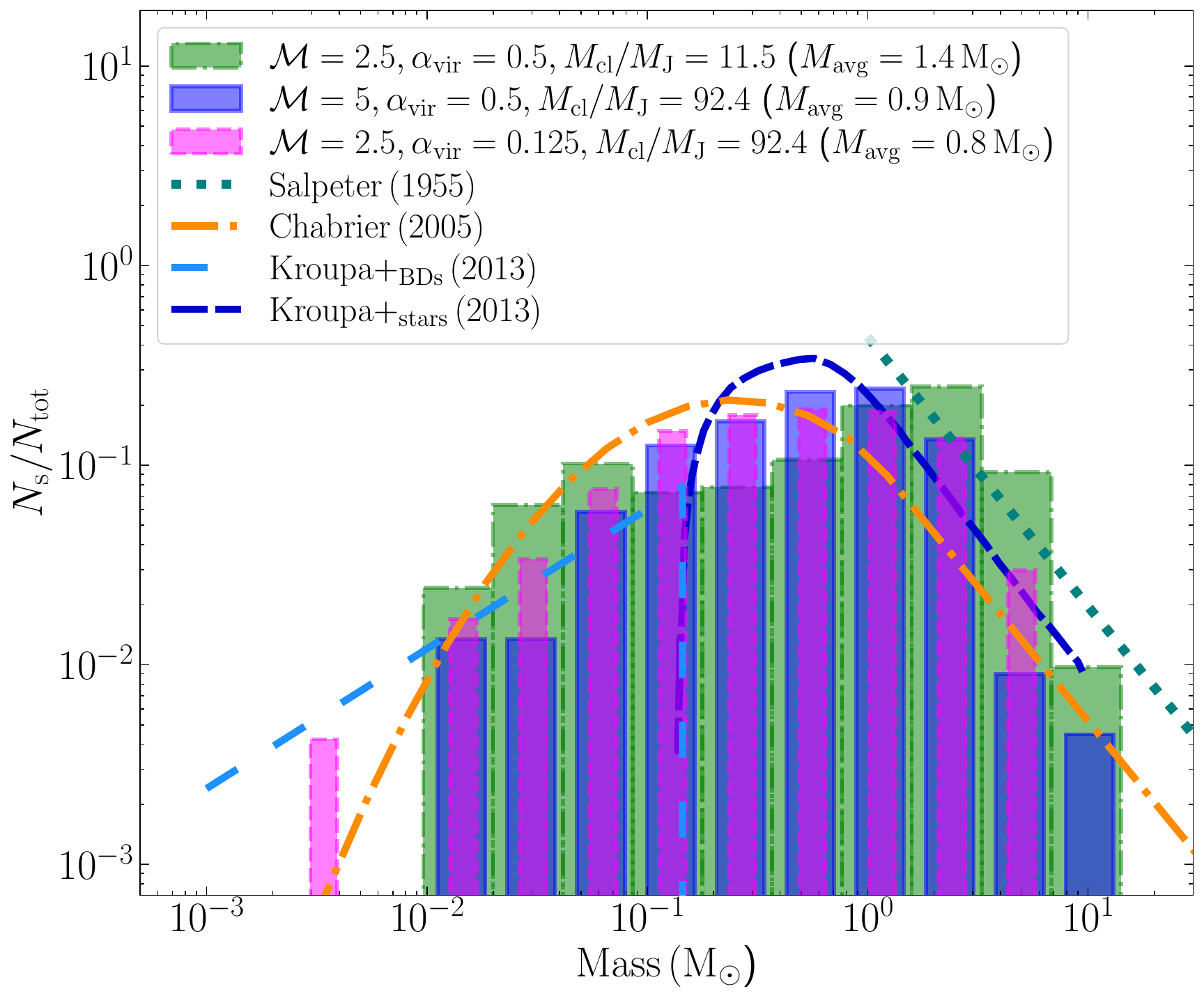}
    \caption{Stellar mass function for simulations with different combinations of $\mathcal{M}, \alpha_{\mathrm{vir}}$, and $M_{\mathrm{cl}}/M_{\mathrm{J}}$ at SFE = 5\%. The green histogram with dash-dotted edges and the blue histogram with solid edges are the same as in Fig.~\ref{fig:imf}. The magenta histogram with dashed edges represents the IMF for the additional simulation model with fixed $M_{\mathrm{cl}}/M_{\mathrm{J}}=92.4$ as in the $\mathcal{M} = 5$ simulation, but $\mathcal{M} = 2.5$ and $\alpha_{\mathrm{vir}}=0.125$. The plotted curves are the same observational IMF models shown in Fig.~\ref{fig:imf}.}  
    \label{fig:imf_comb}
\end{figure}

\subsubsection{On isolating the effects of $\mathcal{M}$, $\alpha_{\mathrm{vir}}$, $M_{\mathrm{cl}}/M_{\mathrm{J}}$}
\label{sec: isolate}
\citet{2016MNRAS.462.4171B} find that increasing the Mach number leads to a higher IMF peak mass in their low-density regime simulations, while the peak remains unchanged in the high-density regime (zoom-in). Neither of these results fully agrees with our findings, primarily due to differences in the underlying numerical setup. \citet{2016MNRAS.462.4171B} increase the Mach number by increasing the velocity dispersion, which in turn raises the cloud virial parameter, since $\alpha_{\mathrm{vir}} \propto \sigma_{\mathrm{v}}^2$. As the virial parameter increases, self-gravity becomes less effective, and only the most massive overdensities are able to overcome internal support and collapse, leading to higher stellar masses \citep{2016MNRAS.462.4171B,mathew2025}. Thus, the increase in $\alpha_{\mathrm{vir}}$ would counteract the effect of the increase in $\mathcal{M}$.

Since it is important to isolate the effect of $\mathcal{M}$ from $\alpha_{\mathrm{vir}}$, we keep $\alpha_{\mathrm{vir}}$ fixed. As a result, the mean thermal Jeans mass $M_\mathrm{J}$ or the number of Jeans masses $M_{\mathrm{cl}}/M_{\mathrm{J}}$ varies across our three simulation models (see Appendix~\ref{sec: isolate_proof}), which could influence the initial fragmentation of the cloud. Therefore, it is possible that the variation in the IMF shape for our $\mathcal{M}=2.5$ simulations (green histogram with dash-dotted edges in Fig.~\ref{fig:imf}) is partly due to the variation in $M_{\mathrm{cl}}/M_{\mathrm{J}}$.

In order to test this, we also carried out an additional set of simulations where we lowered the Mach number from $\mathcal{M}=5$ to $\mathcal{M}=2.5$ by reducing the velocity dispersion $\sigma_\mathrm{v}$, but keeping the mean density the same, which lowers $\alpha_{\mathrm{vir}}$ (and also increases $\mu_{\mathrm{B}}$), but keeps $M_{\mathrm{cl}}/M_{\mathrm{J}}$ fixed. In this experiment, we find that the IMF does not change significantly (see Fig.~\ref{fig:imf_comb}), similar to the findings of \citet{2016MNRAS.462.4171B} \citep[see also][]{2019MNRAS.489.1880H}. This should not be directly interpreted to mean that the Mach number has no effect on the IMF, or that the IMF variations seen in the previous $\mathcal{M}=2.5$ simulations in Fig.~\ref{fig:imf} are solely due to changes in $M_{\mathrm{cl}}/M_{\mathrm{J}}$. This is because, in the additional simulations, although $M_{\mathrm{cl}}/M_{\mathrm{J}}$ was fixed, $\alpha_{\mathrm{vir}}$ decreased (see Tab.~\ref{tab:init}). In relatively low Mach number regimes ($\mathcal{M}\sim2.5$), the effect of increased gravitational dominance (lower $\alpha_{\mathrm{vir}}$) will be significant as the shocks are weaker in dispersing the overdensities. As mentioned above, the increase in fragmentation due to the decrease in $\alpha_{\mathrm{vir}}$ can counteract the decrease in fragmentation due to the lowering of $\mathcal{M}$, which in turn, can result in an almost unchanged IMF.

Also, to keep $\mathcal{M}_{\mathrm{A}}$ fixed, $B_0$ had to be varied in the additional simulations, which increased the dimensionless quantity $\mu_\mathrm{B}$ by a factor of $2$ (see Tab.~\ref{tab:init}). The higher $\mu_\mathrm{B}$ can increase the fragmentation as a result of the reduced magnetic support \citep{2008MNRAS.385.1820P,2011A&A...528A..72H,2014MNRAS.439.3420M,2018MNRAS.476..771C,2019FrASS...6....7K}. This effect is in the opposite direction to that of lowering $\mathcal{M}$, and could therefore diminish the net variation in the IMF. Further, it is shown in \citet{mathew2025} that, at $\mathcal{M} = 5$, when $\alpha_{\mathrm{vir}}$ decreases by a factor of 4 and $M_{\mathrm{cl}}/M_{\mathrm{J}}$ increases by a factor of $8$, the IMF did not change significantly.

Moreover, as star formation begins, the radiative heating by stars raises the local temperature and Jeans mass on sub-parsec scales \citep{bate2009,2011ApJ...740...74K,2016MNRAS.458..673G,2017JPhCS.837a2007F,mathew2020,hennebelle2020}, such that the dependence on the initial mean thermal Jeans mass becomes negligible at the later stages.

Taken together, this means that the IMF shape for the $\mathcal{M}=2.5$ simulations in Fig.~\ref{fig:imf} cannot be explained by just the variation in $M_{\mathrm{cl}}/M_{\mathrm{J}}$ \citep[see also][]{2019A&A...622A.125L}. In addition, changes in the Mach number have been shown to greatly alter the density distribution and fragmentation of the cloud, right from the earliest stages, where only turbulence plays a role \citep[e.g.,][see also Fig.~\ref{fig:densmap1}]{sfr_fk2012}.

Finally, we note it is not possible to fully isolate the effects of $\mathcal{M}$, $\alpha_{\mathrm{vir}}$, and $M_{\mathrm{cl}}/M_{\mathrm{J}}$ individually, as they are inter-related (see Appendix~\ref{sec: isolate_proof}). Consequently, we cannot design a numerical experiment in which one of these parameters is varied between simulations while keeping the other two fixed simultaneously.

\subsection{Resolution study}
\label{sec:res_study}
We carry out a resolution study to test the numerical convergence of our results. Given the computational expense of high-resolution runs, we conduct the convergence analysis for the $\mathcal{M}=2.5$ model as a representative case. The convergence test for the $\mathcal{M}=2.5$ model allows us to verify whether the higher average stellar mass observed in these simulations, as compared to the other $\mathcal{M}$ models, holds at higher resolution levels. For $\mathcal{M}=2.5$, we perform additional simulations at AMR resolution levels two times lower and two times higher than that of the simulations used in this study, i.e., at $\Delta x = 400\, \mathrm{AU}$ and $\Delta x = 25\, \mathrm{AU}$, compared to the base model with a resolution of $\Delta x = 100\, \mathrm{AU}$. The additional simulations use the same turbulence realisation (i.e., the same random seed) as the base model. Tab.~\ref{tab:res_study} lists the number of stars formed and the average stellar mass for the three resolution levels at $\mathrm{SFE}=5\%$. We find that the simulation with the coarsest resolution ($\Delta x = 400\, \mathrm{AU}$) produces only one star, while the base model ($\Delta x = 100\, \mathrm{AU}$) and the higher-resolution model ($\Delta x = 25\, \mathrm{AU}$) produce 12 and 13~star particles, respectively. The single star formed with $\Delta x = 400\, \mathrm{AU}$ has a mass of $9.8\,\msol$, while the average mass is $0.80\,\msol$ and $0.77\,\msol$ for $\Delta x = 100\,\mathrm{AU}$ and $25\,\mathrm{AU}$, respectively, indicating good convergence at our base resolution of $\Delta x = 100\,\mathrm{AU}$.

\begin{table}
	\caption{Resolution study for a single $\mach=2.5$ simulation.}
	\label{tab:res_study}
	\begin{tabular}{c|ccc|} 
	    \hline
		\hline
         & $\Delta x = 25\ \mathrm{AU}$ & $\Delta x = 100\ \mathrm{AU}$ & $\Delta x = 400\ \mathrm{AU}$\\
		\hline
		\hline 
		Number of stars & $13$ & $12$ & $1$\\
		Average stellar mass & $0.77\,\msol$ & $0.80\,\msol$ & $9.8\,\msol$\\
		\hline
	\end{tabular}
	\\
    \raggedright \textbf{Notes.} Number of star particles formed and their average mass at $\mathrm{SFE}=5\%$ for different levels of numerical resolution, quantified by the minimum grid cell length, $\Delta x = 25\ \mathrm{AU}$ (column~2), $\Delta x = 100\ \mathrm{AU}$ (column~3; as in the base model), and $400\ \mathrm{AU}$ (column~4).
\end{table}

\subsection{Multiplicity}
\label{sec: multiplicity}
Our simulations produce a large number of binary systems. To investigate the multiplicity statistics, we identify multiple systems of different orders (singles, binaries, triples, and quadruples) in our simulations based on the algorithm employed in \citet{2009MNRAS.392..590B}. The systems are identified by grouping the closest gravitationally bound objects. The procedure here does not add new members to groups with order higher than quadruples since higher-order systems are generally dynamically unstable and tend to decay to lower-order systems with cloud evolution. For a detailed description of the multiplicity algorithm, see \citet{mathew2021} and \citet{mathew2023}. 

With the list of multiple systems derived using the above algorithm, we compute the multiplicity fraction in different mass ranges. The multiplicity fraction ($mf$) within a given mass range is defined as the ratio of multiple systems to the total number of systems in which the primary star lies within that mass range, i.e.,
\begin{equation}
    \label{eq:mf}
    mf =  \frac{B + T + Q}{S + B + T + Q},
\end{equation}
where $S$, $B$, $T$, and $Q$ represent the number of singles, binaries, triples, and quadruples, respectively.

\begin{figure}
    \centering
    \includegraphics[width=\columnwidth]{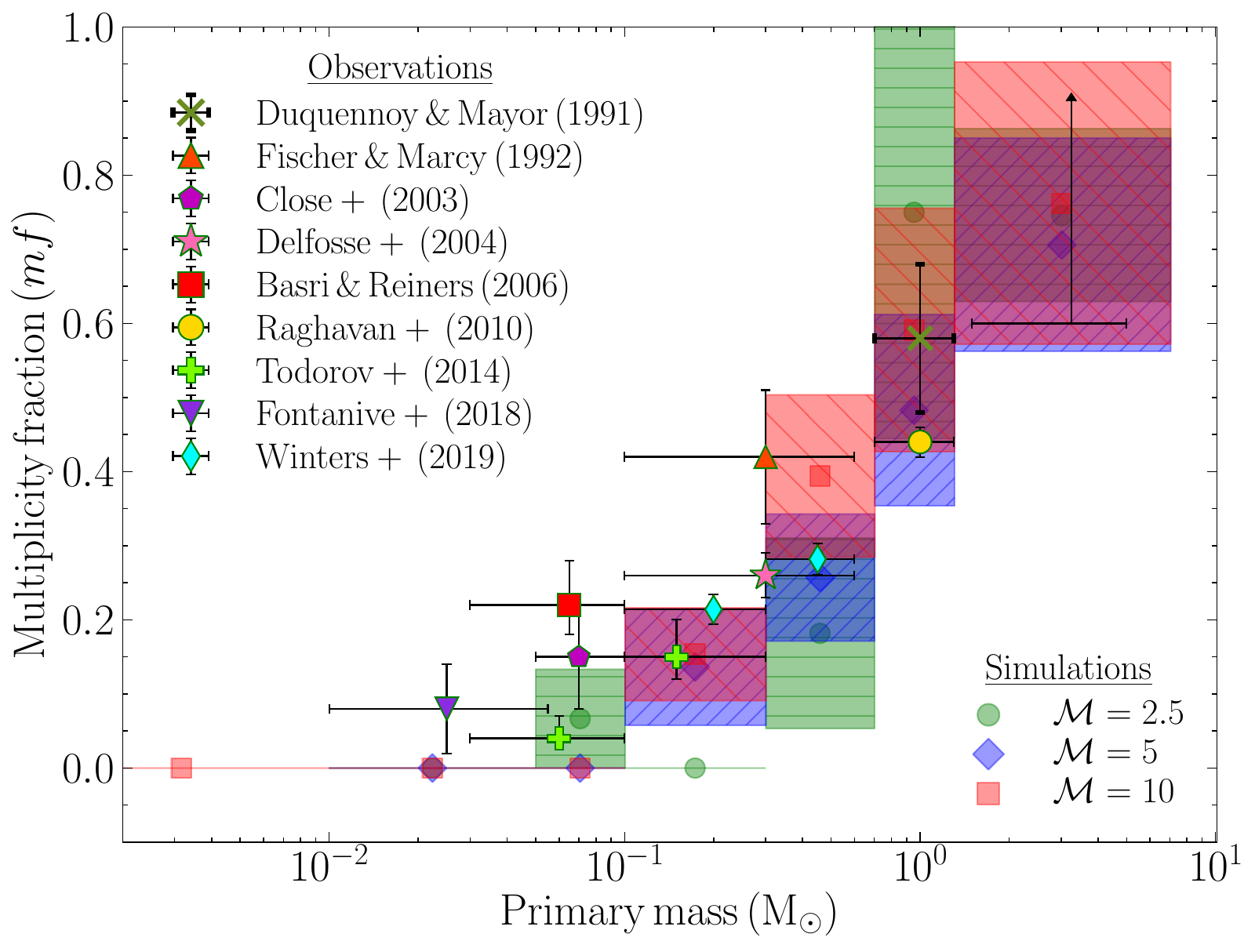}
    \caption{Multiplicity fraction ($mf$) calculated using Eq.~(\ref{eq:mf}) in different primary mass intervals for the simulation models with $\mathcal{M} = 2.5$ (green circles and horizontally hatched boxes), $\mathcal{M} = 5$ (blue diamonds and boxes with right-leaning diagonal hatching) and $\mathcal{M} = 10$ (red squares and boxes with left-leaning diagonal hatching). The markers represent $mf$ in the mass range represented by the width of the box surrounding the marker. The error limit in the $mf$ is denoted by the box height. The data points with error bars show the multiplicity fractions from various observations, with the horizontal and vertical error bars representing the mass interval studied in the survey and the uncertainties, respectively. The observational data (from low to high primary mass) are derived from \citet{2018MNRAS.479.2702F}, \citet{2014ApJ...788...40T}, \citet{2006AJ....132..663B}, \citet{2003ApJ...587..407C}, \citet{2014ApJ...788...40T}, \citet{2019AJ....157..216W} (not corrected for undetected companions), \citet{2004ASPC..318..166D}, \citet{1992ApJ...396..178F}, \citet{2010ApJS..190....1R} and \citet{1991A&A...248..485D}. The $mf$ for high-mass stars is not well understood. The lower limit of $mf$ in the mass range of $1.5$--$5\,\mathrm{M_\odot}$ is $\sim0.5$--$0.6$ \citep{2012MNRAS.424.1925C,2013ARA&A..51..269D}. Massive stars typically have $mf\sim1$ \citep{2009AJ....137.3358M,2011IAUS..272..474S,2017A&A...599L...9S,2020SSRv..216...70L}.}
    \label{fig:mf}
\end{figure}

Fig.~\ref{fig:mf} shows the multiplicity fraction as a function of the primary mass, where the primary is the highest-mass star in the system. All three models show an increasing trend in multiplicity fraction with mass, which is consistent with observations \citep[see, e.g.,][]{2013ARA&A..51..269D} and previous numerical works \citep{2012MNRAS.419.3115B,2012ApJ...754...71K,2018MNRAS.476..771C,2020MNRAS.497..336S,mathew2021,offner2023}. The $mf$ is slightly higher for $\mathcal{M}=10$ than that for $\mathcal{M}=5$ in every mass interval above $0.1\, \mathrm{M_\odot}$. Further, $mf$ for the $\mathcal{M}=2.5$ simulations increases more sharply compared to the other models around the mass range $0.3-1\,\msol$, with $mf$ values much higher than those for $\mathcal{M}=5$ and $10$ above $\sim1\,\msol$. It is also approximately in this range where the valley of the IMF for $\mathcal{M}=2.5$ is located, suggesting that the signatures of bimodality can also be observed in the multiplicity characteristics. 

\subsection{Specific angular momentum}
\label{sec:angmom}
Observations suggest that the angular momentum decreases significantly during the star formation process. The specific angular momentum ($j$) of dense molecular cloud cores (diameter $\sim\!0.1\,$pc) is found to be greater than $10^{21}\, \mathrm{cm^2\, s^{-1}}$ \citep{1993ApJ...406..528G,2000ApJ...543..822B,2002ApJ...572..238C}. As the collapse proceeds, this value is reduced to between $10^{17}$--$10^{21}\, \mathrm{cm^2\, s^{-1}}$ in class~0/I envelopes and binary systems \citep{1992ASPC...32...41S,1997ApJ...488..317O,2015ApJ...799..193Y,2020A&A...637A..92G}, and further drops to $10^{16}$--$10^{17}\, \mathrm{cm^2\, s^{-1}}$ in T-Tauri stars \citep{1986ApJ...309..275H}.

\begin{figure}
    \centering
    \includegraphics[width=\columnwidth]{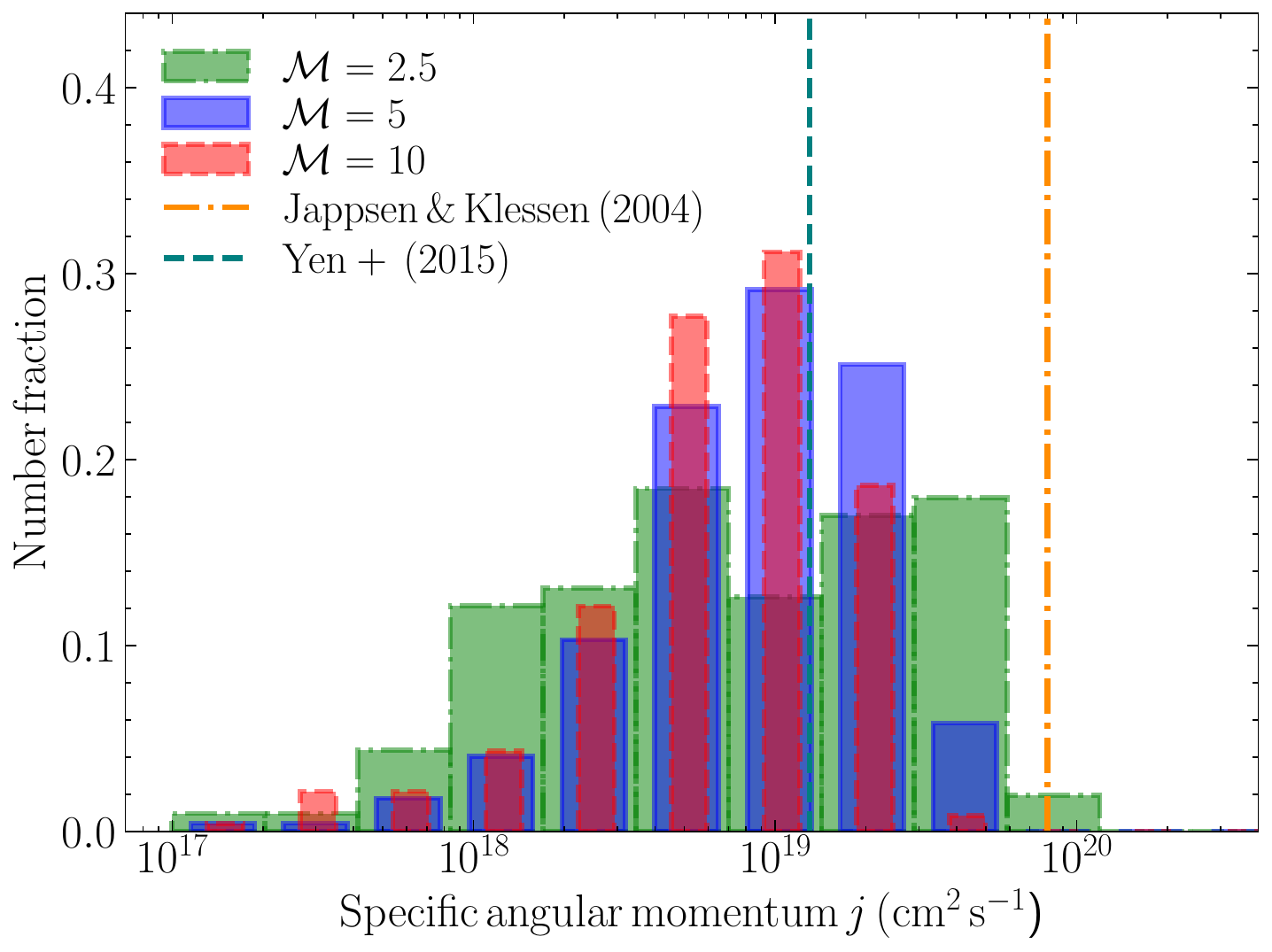}
    \caption{Distribution of the specific angular momentum ($j$) of the star particles in the simulations with $\mathcal{M}=2.5$ (green histogram with dash-dotted edges), $\mathcal{M}=5$ (blue histogram with solid edges), and $\mathcal{M}=10$ (red histogram with dashed edges). The vertical dashed line represents the $j$ value from observations for the class~0 protostar B335 at $\sim\!$~180~AU by \citet{2015ApJ...812..129Y}, and the vertical dash-dotted line corresponds to the mean $j$ measured in the simulations of \citet{2004A&A...423....1J} with a star particle radius of 560~AU.}
    \label{fig:angmom}
\end{figure}

The specific angular momentum distribution of the star particles that form in our  three main simulation models is shown in Fig.~\ref{fig:angmom}. The range of specific angular momentum of the star particles ($\sim\!10^{17}$--$10^{20}\, \mathrm{cm^2\, s^{-1}}$) spans the regime of protostellar envelopes and binaries, with a few of them having $j$ values typical of T-Tauri stars. The average specific angular momentum is slightly higher for $\mathcal{M}=2.5$ ($j_{\mathrm{mean}} = 1.5 \times 10^{19}\, \mathrm{cm^2\, s^{-1}}$) than for $\mathcal{M}=10$ ($j_{\mathrm{mean}} = 8.8 \times 10^{18}\, \mathrm{cm^2\, s^{-1}}$). The $j_{\mathrm{mean}}$ for $\mathcal{M}=5$ is in between, at $1.2 \times 10^{19}\, \mathrm{cm^2\, s^{-1}}$. This suggests that the specific angular momentum of stars is weakly dependent on the large-scale turbulent Mach number of the host cloud. Further, the distribution for $\mathcal{M}=2.5$ is bimodal with the valley occurring around the range in which the peak of the distributions for the higher Mach numbers exists, similar to the trend seen in the case of the IMF. This implies that the stellar angular momentum distribution is closely linked to the IMF.

\section{Conclusions}
\label{sec:conclusions}
We carried out a suite of hydrodynamical simulations of the collapse of molecular clouds, including gravity, turbulence, magnetic fields, and feedback in the form of jets and outflows and accretion heating, to study the effect of the cloud's turbulent Mach number on the star cluster formation process. The set consists of three different simulation models with Mach numbers $\mathcal{M}=2.5, 5$, and $10$. The simulations in each of the models are evolved until a star formation efficiency of 5\%. To isolate the effect of $\mach$ from fundamental physical quantities such as the virial parameter $\alpha_{\mathrm{vir}}$ and the Alfv\'en Mach number $\mathcal{M_\mathrm{A}}$, we ensure that they remain the same in the three models. As a consequence of this choice, the number of Jeans masses $M_{\mathrm{cl}}/M_{\mathrm{J}}$ varies between the simulation models, which could also influence the results. However, fully isolating the effects of $\mathcal{M}$ from $\alpha_{\mathrm{vir}}$ and $M_{\mathrm{cl}}/M_{\mathrm{J}}$ simultaneously is not possible (see \S\ref{sec: isolate} and Appendix~\ref{sec: isolate_proof}). 

We find that the star formation rate and the initial mass function have a non-linear dependence on the Mach number. For $\mathcal{M}=10$, we see a higher degree of fragmentation (more stars) and higher accretion rates compared to the $\mathcal{M}=2.5$ and $5$ cases. Consequently, for $\mathcal{M}=10$, the star formation rate per freefall time ($\mathrm{SFR_{ff}}$) is almost 3~times that of the other two models. While the $\mathcal{M}=5$ model produces a higher number of stars overall as compared to the $\mathcal{M}=2.5$ case, the mean accretion rate and $\mathrm{SFR_{ff}}$ become comparable for both cases by $\mathrm{SFE\sim3\%}$. Further, we find that the IMF obtained in the $\mathcal{M}=5$ and $10$ models exhibits similar forms, although the IMF peak for $\mathcal{M}=10$ is slightly lower. On the other hand, the shape of the IMF for $\mathcal{M}=2.5$ is significantly different from the higher-$\mach$ models. The IMF for $\mathcal{M}=2.5$ is bimodal with a primary peak at supersolar masses and a secondary peak around the substellar regime.

The (primary) peak for $\mathcal{M}=2.5$ is at a higher mass than the peak of the (uni-modal) IMFs for $\mathcal{M}=5$ and $\mathcal{M}=10$, and also, we find that the average stellar mass decreases with increasing $\mathcal{M}$, both of which are consistent with gravo-turbulent theories of star formation \citep{2002ApJ...576..870P, 2008ApJ...684..395H,2009ApJ...702.1428H,2012MNRAS.423.2037H,2013MNRAS.430.1653H,2024ARA&A..62...63H}. Since turbulent shocks are weaker for low Mach numbers, over-densities (analogous to dense cores) forming in the shocked regions are less dense. As a consequence, the Jeans mass is higher and the fragmentation is lower, leading to higher-mass stars forming for lower $\mach$ compared to higher $\mach$ \citep{2002ApJ...576..870P,2008ApJ...684..395H,2012MNRAS.423.2037H,2016MNRAS.462.4171B,Haugb_lle_2018}. This is why we find that the primary IMF peak and the average stellar mass are higher in the case of $\mathcal{M}=2.5$ than at higher $\mach$. However, it is to be noted that the existence of a secondary peak in the $\mathcal{M}=2.5$ model remains unexplained within the framework of the current theories of the IMF, pointing to the need for their refinement. The connection between the bimodality observed in this study and the origin of the IMF will be addressed in detail in a follow-up work.

We also show that the multiplicity fraction ($mf$) is influenced by the turbulent Mach number of the cloud. While the multiplicity fraction is relatively lower for $\mathcal{M}=2.5$ at low masses, it shows a sharper increase in $mf$ with mass compared to the cases with higher Mach numbers. We also find a weak negative correlation between the mean specific angular momentum of the stars ($j_{\mathrm{mean}}$) and the cloud Mach number, with the $\mathcal{M}=2.5$ simulations having the highest $j_{\mathrm{mean}}$. Further, the angular momentum distribution for $\mathcal{M}=2.5$ is bimodal like its IMF, suggesting a close link between the stellar angular momentum and mass.

Overall, using several results, we demonstrate that the turbulent Mach number of the cloud significantly alters the properties of star formation and the IMF, especially at low Mach numbers and this must be considered in future theory, simulations, and observations of star cluster formation.

\section*{Acknowledgements}
We thank the anonymous referee whose comments and suggestions improved the quality of this work. C.~F.~acknowledges funding provided by the Australian Research Council (Discovery Projects DP230102280 and DP250101526), and the Australia-Germany Joint Research Cooperation Scheme (UA-DAAD). A.~S.~acknowledges support from the Australian Research Council's Discovery Early Career Researcher Award (DECRA, project~DE250100003) and the Australia-Germany Joint Research Cooperation Scheme of Universities Australia (UA--DAAD, 2025--2026). We further acknowledge high-performance computing resources provided by the Leibniz Rechenzentrum and the Gauss Centre for Supercomputing (grants~pr32lo, pr48pi and GCS Large-scale project~10391), the Australian National Computational Infrastructure (grant~ek9) and the Pawsey Supercomputing Centre (project~pawsey0810) in the framework of the National Computational Merit Allocation Scheme and the ANU Merit Allocation Scheme. The simulation software FLASH was in part developed by the DOE-supported Flash Center for Computational Science at the University of Chicago and the University of Rochester. The turbulence driving implemented in the simulations utilises the publicly available {\sc TurbGen} code \citep{2010A&A...512A..81F,2022ascl.soft04001F}. This work makes use of the yt-project \citep{yt} and colourmaps in the CMasher package \citep{cmasher}. 

\section*{Data Availability}
The data and codes used in this article are available upon reasonable request to the authors.




\bibliographystyle{mnras}
\bibliography{Bibliography} 

@ARTICLE{2008MNRAS.385.1820P,
       author = {{Price}, Daniel J. and {Bate}, Matthew R.},
        title = "{The effect of magnetic fields on star cluster formation}",
      journal = {\mnras},
         year = 2008,
        month = apr,
       volume = {385},
       number = {4},
        pages = {1820-1834},
          doi = {10.1111/j.1365-2966.2008.12976.x},
archivePrefix = {arXiv},
       eprint = {0801.3293},
 primaryClass = {astro-ph},
       adsurl = {https://ui.adsabs.harvard.edu/abs/2008MNRAS.385.1820P}
}

@ARTICLE{2007ARA&A..45..481Z,
       author = {{Zinnecker}, Hans and {Yorke}, Harold W.},
        title = "{Toward Understanding Massive Star Formation}",
      journal = {\araa},
         year = 2007,
        month = sep,
       volume = {45},
       number = {1},
        pages = {481-563},
          doi = {10.1146/annurev.astro.44.051905.092549},
archivePrefix = {arXiv},
       eprint = {0707.1279},
 primaryClass = {astro-ph},
       adsurl = {https://ui.adsabs.harvard.edu/abs/2007ARA&A..45..481Z}
}

@ARTICLE{2011A&A...528A..72H,
       author = {{Hennebelle}, P. and {Commer{\c{c}}on}, B. and {Joos}, M. and {Klessen}, R.~S. and {Krumholz}, M. and {Tan}, J.~C. and {Teyssier}, R.},
        title = "{Collapse, outflows and fragmentation of massive, turbulent and magnetized prestellar barotropic cores}",
      journal = {\aap},
         year = 2011,
        month = apr,
       volume = {528},
          eid = {A72},
        pages = {A72},
          doi = {10.1051/0004-6361/201016052},
archivePrefix = {arXiv},
       eprint = {1101.1574},
 primaryClass = {astro-ph.GA},
       adsurl = {https://ui.adsabs.harvard.edu/abs/2011A&A...528A..72H}
}

@INPROCEEDINGS{2017JPhCS.837a2007F,
       author = {{Federrath}, Christoph and {Krumholz}, Mark and {Hopkins}, Philip F.},
        title = "{Converging on the Initial Mass Function of Stars}",
    booktitle = {Journal of Physics Conference Series},
         year = "2017",
       volume = {837},
        month = "May",
          eid = {012007},
        pages = {012007},
          doi = {10.1088/1742-6596/837/1/012007},
       adsurl = {https://ui.adsabs.harvard.edu/abs/2017JPhCS.837a2007F}
}

@ARTICLE{2019MNRAS.489.1880H,
       author = {{He}, Chong-Chong and {Ricotti}, Massimo and {Geen}, Sam},
        title = "{Simulating star clusters across cosmic time - I. Initial mass function, star formation rates, and efficiencies}",
      journal = {\mnras},
         year = 2019,
        month = oct,
       volume = {489},
       number = {2},
        pages = {1880-1898},
          doi = {10.1093/mnras/stz2239},
archivePrefix = {arXiv},
       eprint = {1904.07889},
 primaryClass = {astro-ph.GA},
       adsurl = {https://ui.adsabs.harvard.edu/abs/2019MNRAS.489.1880H}
}

@ARTICLE{2013MNRAS.435.1701C,
       author = {{Col{\'\i}n}, Pedro and {V{\'a}zquez-Semadeni}, Enrique and {G{\'o}mez}, Gilberto C.},
        title = "{Molecular cloud evolution - V. Cloud destruction by stellar feedback}",
      journal = {\mnras},
         year = 2013,
        month = oct,
       volume = {435},
       number = {2},
        pages = {1701-1714},
          doi = {10.1093/mnras/stt1409},
archivePrefix = {arXiv},
       eprint = {1304.1570},
 primaryClass = {astro-ph.GA},
       adsurl = {https://ui.adsabs.harvard.edu/abs/2013MNRAS.435.1701C}
}

@ARTICLE{2000ApJS..131..273F,
       author = {{Fryxell}, B. and {Olson}, K. and {Ricker}, P. and {Timmes}, F.~X. and
         {Zingale}, M. and {Lamb}, D.~Q. and {MacNeice}, P. and {Rosner}, R. and
         {Truran}, J.~W. and {Tufo}, H.},
        title = "{FLASH: An Adaptive Mesh Hydrodynamics Code for Modeling Astrophysical Thermonuclear Flashes}",
      journal = {\apjs},
         year = "2000",
        month = "Nov",
       volume = {131},
       number = {1},
        pages = {273-334},
          doi = {10.1086/317361},
       adsurl = {https://ui.adsabs.harvard.edu/abs/2000ApJS..131..273F}
}

@ARTICLE{2025arXiv251012203F,
       author = {{Federrath}, Christoph and {Offner}, Stella},
        title = "{Computational advances and challenges in simulations of turbulence and star formation}",
      journal = {arXiv e-prints},
         year = 2025,
        month = oct,
          eid = {arXiv:2510.12203},
        pages = {arXiv:2510.12203},
          doi = {10.48550/arXiv.2510.12203},
archivePrefix = {arXiv},
       eprint = {2510.12203},
 primaryClass = {astro-ph.GA},
       adsurl = {https://ui.adsabs.harvard.edu/abs/2025arXiv251012203F}
}

@ARTICLE{2021ApJ...911..128K,
       author = {{Kim}, Jeong-Gyu and {Ostriker}, Eve C. and {Filippova}, Nina},
        title = "{Star Formation Efficiency and Dispersal of Giant Molecular Clouds with UV Radiation Feedback: Dependence on Gravitational Boundedness and Magnetic Fields}",
      journal = {\apj},
         year = 2021,
        month = apr,
       volume = {911},
       number = {2},
          eid = {128},
        pages = {128},
          doi = {10.3847/1538-4357/abe934},
archivePrefix = {arXiv},
       eprint = {2011.07772},
 primaryClass = {astro-ph.GA},
       adsurl = {https://ui.adsabs.harvard.edu/abs/2021ApJ...911..128K}
}

@ARTICLE{2008MNRAS.386....3C,
       author = {{Clark}, Paul C. and {Bonnell}, Ian A. and {Klessen}, Ralf S.},
        title = "{The star formation efficiency and its relation to variations in the initial mass function}",
      journal = {\mnras},
         year = 2008,
        month = may,
       volume = {386},
       number = {1},
        pages = {3-10},
          doi = {10.1111/j.1365-2966.2008.13005.x},
archivePrefix = {arXiv},
       eprint = {0803.4053},
 primaryClass = {astro-ph},
       adsurl = {https://ui.adsabs.harvard.edu/abs/2008MNRAS.386....3C}
}

@ARTICLE{2006ApJ...637..384B,
       author = {{Ballesteros-Paredes}, Javier and {Gazol}, Adriana and {Kim}, Jongsoo and {Klessen}, Ralf S. and {Jappsen}, Anne-Katharina and {Tejero}, Epimenio},
        title = "{The Mass Spectra of Cores in Turbulent Molecular Clouds and Implications for the Initial Mass Function}",
      journal = {\apj},
         year = 2006,
        month = jan,
       volume = {637},
       number = {1},
        pages = {384-391},
          doi = {10.1086/498228},
archivePrefix = {arXiv},
       eprint = {astro-ph/0509591},
 primaryClass = {astro-ph},
       adsurl = {https://ui.adsabs.harvard.edu/abs/2006ApJ...637..384B}
}

@ARTICLE{2005ApJ...630L..45K,
       author = {{Kim}, Jongsoo and {Ryu}, Dongsu},
        title = "{Density Power Spectrum of Compressible Hydrodynamic Turbulent Flows}",
      journal = {\apjl},
         year = 2005,
        month = sep,
       volume = {630},
       number = {1},
        pages = {L45-L48},
          doi = {10.1086/491600},
archivePrefix = {arXiv},
       eprint = {astro-ph/0507591},
 primaryClass = {astro-ph},
       adsurl = {https://ui.adsabs.harvard.edu/abs/2005ApJ...630L..45K}
}

@INPROCEEDINGS{2008ASPC..385..145D,
       author = {{Dubey}, A. and {Fisher}, R. and {Graziani}, C. and {Jordan}, G.~C., IV and
         {Lamb}, D.~Q. and {Reid}, L.~B. and {Rich}, P. and {Sheeler}, D. and
         {Townsley}, D. and {Weide}, K.},
        title = "{Challenges of Extreme Computing using the FLASH code}",
    booktitle = {Numerical Modeling of Space Plasma Flows},
         year = "2008",
       editor = {{Pogorelov}, N.~V. and {Audit}, E. and {Zank}, G.~P.},
       series = {Astronomical Society of the Pacific Conference Series},
       volume = {385},
        month = "Apr",
        pages = {145},
       adsurl = {https://ui.adsabs.harvard.edu/abs/2008ASPC..385..145D}
}

@ARTICLE{2010A&A...512A..81F,
       author = {{Federrath}, C. and {Roman-Duval}, J. and {Klessen}, R.~S. and
         {Schmidt}, W. and {Mac Low}, M. -M.},
        title = "{Comparing the statistics of interstellar turbulence in simulations and observations. Solenoidal versus compressive turbulence forcing}",
      journal = {\aap},
         year = "2010",
        month = "Mar",
       volume = {512},
          eid = {A81},
        pages = {A81},
          doi = {10.1051/0004-6361/200912437},
archivePrefix = {arXiv},
       eprint = {0905.1060},
 primaryClass = {astro-ph.SR},
       adsurl = {https://ui.adsabs.harvard.edu/abs/2010A&A...512A..81F}
}

@ARTICLE{2010ApJ...713..269F,
       author = {{Federrath}, Christoph and {Banerjee}, Robi and {Clark}, Paul C. and
         {Klessen}, Ralf S.},
        title = "{Modeling Collapse and Accretion in Turbulent Gas Clouds: Implementation and Comparison of Sink Particles in AMR and SPH}",
      journal = {\apj},
         year = "2010",
        month = "Apr",
       volume = {713},
       number = {1},
        pages = {269-290},
          doi = {10.1088/0004-637X/713/1/269},
       archivePrefix = {arXiv},
       eprint = {1001.4456},
 primaryClass = {astro-ph.SR},
       adsurl = {https://ui.adsabs.harvard.edu/abs/2010ApJ...713..269F}
}

@ARTICLE{2009ApJ...703..131O,
       author = {{Offner}, Stella S.~R. and {Klein}, Richard I. and
         {McKee}, Christopher F. and {Krumholz}, Mark R.},
        title = "{The Effects of Radiative Transfer on Low-Mass Star Formation}",
      journal = {\apj},
         year = "2009",
        month = "Sep",
       volume = {703},
       number = {1},
        pages = {131-149},
          doi = {10.1088/0004-637X/703/1/131},
archivePrefix = {arXiv},
       eprint = {0904.2004},
 primaryClass = {astro-ph.SR},
       adsurl = {https://ui.adsabs.harvard.edu/abs/2009ApJ...703..131O}
}

@ARTICLE{sfr_fk2012,
       author = {{Federrath}, Christoph and {Klessen}, Ralf S.},
        title = "{The Star Formation Rate of Turbulent Magnetized Clouds: Comparing Theory, Simulations, and Observations}",
      journal = {\apj},
         year = "2012",
        month = "Dec",
       volume = {761},
       number = {2},
          eid = {156},
        pages = {156},
          doi = {10.1088/0004-637X/761/2/156},
archivePrefix = {arXiv},
       eprint = {1209.2856},
 primaryClass = {astro-ph.SR},
       adsurl = {https://ui.adsabs.harvard.edu/abs/2012ApJ...761..156F}
}

@ARTICLE{2019FrASS...6....7K,
       author = {{Krumholz}, Mark R. and {Federrath}, Christoph},
        title = "{The Role of Magnetic Fields in Setting the Star Formation Rate and the Initial Mass Function}",
      journal = {Frontiers in Astronomy and Space Sciences},
         year = "2019",
        month = "Feb",
       volume = {6},
          eid = {7},
        pages = {7},
          doi = {10.3389/fspas.2019.00007},
archivePrefix = {arXiv},
       eprint = {1902.02557},
 primaryClass = {astro-ph.GA},
       adsurl = {https://ui.adsabs.harvard.edu/abs/2019FrASS...6....7K}
}

@ARTICLE{2024JApA...45...17S,
       author = {{Soam}, Archana and {Eswaraiah}, Chakali and {Seta}, Amit and {Dewangan}, Lokesh and {Maheswar}, G.},
        title = "{Turbulence and magnetic fields in star formation}",
      journal = {Journal of Astrophysics and Astronomy},
         year = 2024,
        month = jun,
       volume = {45},
       number = {1},
          eid = {17},
        pages = {17},
          doi = {10.1007/s12036-024-10005-z},
archivePrefix = {arXiv},
       eprint = {2402.18840},
 primaryClass = {astro-ph.GA},
       adsurl = {https://ui.adsabs.harvard.edu/abs/2024JApA...45...17S}
}

@ARTICLE{2004RvMP...76..125M,
       author = {{Mac Low}, Mordecai-Mark and {Klessen}, Ralf S.},
        title = "{Control of star formation by supersonic turbulence}",
      journal = {Reviews of Modern Physics},
         year = "2004",
        month = "Jan",
       volume = {76},
       number = {1},
        pages = {125-194},
          doi = {10.1103/RevModPhys.76.125},
archivePrefix = {arXiv},
       eprint = {astro-ph/0301093},
 primaryClass = {astro-ph},
       adsurl = {https://ui.adsabs.harvard.edu/abs/2004RvMP...76..125M}
}

@ARTICLE{2014ApJ...790..128F,
       author = {{Federrath}, Christoph and {Schr{\"o}n}, Martin and {Banerjee}, Robi and
         {Klessen}, Ralf S.},
        title = "{Modeling Jet and Outflow Feedback during Star Cluster Formation}",
      journal = {\apj},
         year = "2014",
        month = "Aug",
       volume = {790},
       number = {2},
          eid = {128},
        pages = {128},
          doi = {10.1088/0004-637X/790/2/128},
archivePrefix = {arXiv},
       eprint = {1406.3625},
 primaryClass = {astro-ph.SR},
       adsurl = {https://ui.adsabs.harvard.edu/abs/2014ApJ...790..128F}
}

@ARTICLE{2000ApJ...531..350M,
       author = {{Masunaga}, Hirohiko and {Inutsuka}, Shu-ichiro},
        title = "{A Radiation Hydrodynamic Model for Protostellar Collapse. II. The Second Collapse and the Birth of a Protostar}",
      journal = {\apj},
         year = "2000",
        month = "Mar",
       volume = {531},
       number = {1},
        pages = {350-365},
          doi = {10.1086/308439},
       adsurl = {https://ui.adsabs.harvard.edu/abs/2000ApJ...531..350M}
}

@ARTICLE{1969MNRAS.145..271L,
       author = {{Larson}, Richard B.},
        title = "{Numerical calculations of the dynamics of collapsing proto-star}",
      journal = {\mnras},
         year = "1969",
        month = "Jan",
       volume = {145},
        pages = {271},
          doi = {10.1093/mnras/145.3.271},
       adsurl = {https://ui.adsabs.harvard.edu/abs/1969MNRAS.145..271L}
}

@ARTICLE{1993ApJ...411..274Y,
       author = {{Yorke}, Harold W. and {Bodenheimer}, Peter and {Laughlin}, Gregory},
        title = "{The Formation of Protostellar Disks. I. 1 M sub sun}",
      journal = {\apj},
         year = "1993",
        month = "Jul",
       volume = {411},
        pages = {274},
          doi = {10.1086/172827},
       adsurl = {https://ui.adsabs.harvard.edu/abs/1993ApJ...411..274Y}
}

@ARTICLE{1955ApJ...121..161S,
       author = {{Salpeter}, Edwin E.},
        title = "{The Luminosity Function and Stellar Evolution.}",
      journal = {\apj},
         year = "1955",
        month = "Jan",
       volume = {121},
        pages = {161},
          doi = {10.1086/145971},
       adsurl = {https://ui.adsabs.harvard.edu/abs/1955ApJ...121..161S}
}

@INPROCEEDINGS{chabrier2005,
       author = {{Chabrier}, Gilles},
        title = "{The Initial Mass Function: From Salpeter 1955 to 2005}",
    booktitle = {The Initial Mass Function 50 Years Later},
         year = "2005",
       editor = {{Corbelli}, E. and {Palla}, F. and {Zinnecker}, H.},
       series = {Astrophysics and Space Science Library},
       volume = {327},
        month = "Jan",
        pages = {41},
          doi = {10.1007/978-1-4020-3407-7_5},
archivePrefix = {arXiv},
       eprint = {astro-ph/0409465},
 primaryClass = {astro-ph},
       adsurl = {https://ui.adsabs.harvard.edu/abs/2005ASSL..327...41C}
}

@ARTICLE{sfr_pn2011,
       author = {{Padoan}, Paolo and {Nordlund}, {\r{A}}ke},
        title = "{The Star Formation Rate of Supersonic Magnetohydrodynamic Turbulence}",
      journal = {\apj},
         year = "2011",
        month = "Mar",
       volume = {730},
       number = {1},
          eid = {40},
        pages = {40},
          doi = {10.1088/0004-637X/730/1/40},
archivePrefix = {arXiv},
       eprint = {0907.0248},
 primaryClass = {astro-ph.GA},
       adsurl = {https://ui.adsabs.harvard.edu/abs/2011ApJ...730...40P}
}

@ARTICLE{2016JPlPh..82f5301F,
       author = {{Federrath}, Christoph},
        title = "{Magnetic field amplification in turbulent astrophysical plasmas}",
      journal = {Journal of Plasma Physics},
         year = "2016",
        month = "Dec",
       volume = {82},
       number = {6},
          eid = {535820601},
        pages = {535820601},
          doi = {10.1017/S0022377816001069},
archivePrefix = {arXiv},
       eprint = {1610.08132},
 primaryClass = {physics.plasm-ph},
       adsurl = {https://ui.adsabs.harvard.edu/abs/2016JPlPh..82f5301F}
}

@ARTICLE{1992A&A...257..715F,
       author = {{Falgarone}, E. and {Puget}, J. -L. and {Perault}, M.},
        title = "{The small-scale density and velocity structure of quiescent molecular clouds.}",
      journal = {\aap},
         year = "1992",
        month = "Apr",
       volume = {257},
        pages = {715-730},
       adsurl = {https://ui.adsabs.harvard.edu/abs/1992A&A...257..715F}
}

@ARTICLE{2008ApJ...681..365E,
       author = {{Elmegreen}, Bruce G. and {Klessen}, Ralf S. and {Wilson}, Christine D.},
        title = "{On the Constancy of the Characteristic Mass of Young Stars}",
      journal = {\apj},
         year = "2008",
        month = "Jul",
       volume = {681},
       number = {1},
        pages = {365-374},
          doi = {10.1086/588725},
archivePrefix = {arXiv},
       eprint = {0803.4411},
 primaryClass = {astro-ph},
       adsurl = {https://ui.adsabs.harvard.edu/abs/2008ApJ...681..365E}
}

@ARTICLE{2003PASP..115..763C,
       author = {{Chabrier}, Gilles},
        title = "{Galactic Stellar and Substellar Initial Mass Function}",
      journal = {\pasp},
         year = "2003",
        month = "Jul",
       volume = {115},
       number = {809},
        pages = {763-795},
          doi = {10.1086/376392},
archivePrefix = {arXiv},
       eprint = {astro-ph/0304382},
 primaryClass = {astro-ph},
       adsurl = {https://ui.adsabs.harvard.edu/abs/2003PASP..115..763C}
}

@ARTICLE{2001MNRAS.322..231K,
       author = {{Kroupa}, Pavel},
        title = "{On the variation of the initial mass function}",
      journal = {\mnras},
         year = "2001",
        month = "Apr",
       volume = {322},
       number = {2},
        pages = {231-246},
          doi = {10.1046/j.1365-8711.2001.04022.x},
archivePrefix = {arXiv},
       eprint = {astro-ph/0009005},
 primaryClass = {astro-ph},
       adsurl = {https://ui.adsabs.harvard.edu/abs/2001MNRAS.322..231K}
}

@ARTICLE{2014MNRAS.439.3420M,
       author = {{Myers}, Andrew T. and {Klein}, Richard I. and {Krumholz}, Mark R. and
         {McKee}, Christopher F.},
        title = "{Star cluster formation in turbulent, magnetized dense clumps with radiative and outflow feedback}",
      journal = {\mnras},
         year = "2014",
        month = "Apr",
       volume = {439},
       number = {4},
        pages = {3420-3438},
          doi = {10.1093/mnras/stu190},
archivePrefix = {arXiv},
       eprint = {1401.6096},
 primaryClass = {astro-ph.GA},
       adsurl = {https://ui.adsabs.harvard.edu/abs/2014MNRAS.439.3420M}
}

@article{Li_2010,
	doi = {10.1088/2041-8205/720/1/l26},
	url = {https://doi.org/10.1088%2F2041-8205%2F720%2F1%2Fl26},
	year = 2010,
	month = {aug},
	publisher = {{IOP} Publishing},
	volume = {720},
	number = {1},
	pages = {L26--L30},
	author = {Zhi-Yun Li and Peng Wang and Tom Abel and Fumitaka Nakamura},
	title = {{LOWERING} {THE} {CHARACTERISTIC} {MASS} {OF} {CLUSTER} {STARS} {BY} {MAGNETIC} {FIELDS} {AND} {OUTFLOW} {FEEDBACK}},
	journal = {The Astrophysical Journal}
}

@ARTICLE{2008ApJ...688L..79F,
       author = {{Federrath}, Christoph and {Klessen}, Ralf S. and {Schmidt}, Wolfram},
        title = "{The Density Probability Distribution in Compressible Isothermal Turbulence: Solenoidal versus Compressive Forcing}",
      journal = {\apjl},
         year = "2008",
        month = "Dec",
       volume = {688},
       number = {2},
        pages = {L79},
          doi = {10.1086/595280},
archivePrefix = {arXiv},
       eprint = {0808.0605},
 primaryClass = {astro-ph},
       adsurl = {https://ui.adsabs.harvard.edu/abs/2008ApJ...688L..79F}
}

@ARTICLE{2012MNRAS.423.2680M,
       author = {{Molina}, F.~Z. and {Glover}, S.~C.~O. and {Federrath}, C. and
         {Klessen}, R.~S.},
        title = "{The density variance-Mach number relation in supersonic turbulence - I. Isothermal, magnetized gas}",
      journal = {\mnras},
         year = "2012",
        month = "Jul",
       volume = {423},
       number = {3},
        pages = {2680-2689},
          doi = {10.1111/j.1365-2966.2012.21075.x},
       adsurl = {https://ui.adsabs.harvard.edu/abs/2012MNRAS.423.2680M}
}

@ARTICLE{2018MNRAS.476..771C,
       author = {{Cunningham}, Andrew J. and {Krumholz}, Mark R. and
         {McKee}, Christopher F. and {Klein}, Richard I.},
        title = "{The effects of magnetic fields and protostellar feedback on low-mass cluster formation}",
      journal = {\mnras},
         year = 2018,
        month = may,
       volume = {476},
       number = {1},
        pages = {771-792},
          doi = {10.1093/mnras/sty154},
archivePrefix = {arXiv},
       eprint = {1709.01277},
 primaryClass = {astro-ph.GA},
       adsurl = {https://ui.adsabs.harvard.edu/abs/2018MNRAS.476..771C}
}

@ARTICLE{2011ApJ...740...74K,
       author = {{Krumholz}, Mark R. and {Klein}, Richard I. and {McKee}, Christopher F.},
        title = "{Radiation-hydrodynamic Simulations of the Formation of Orion-like Star Clusters. I. Implications for the Origin of the Initial Mass Function}",
      journal = {\apj},
         year = 2011,
        month = oct,
       volume = {740},
       number = {2},
          eid = {74},
        pages = {74},
          doi = {10.1088/0004-637X/740/2/74},
archivePrefix = {arXiv},
       eprint = {1104.2038},
 primaryClass = {astro-ph.GA},
       adsurl = {https://ui.adsabs.harvard.edu/abs/2011ApJ...740...74K}
}

@ARTICLE{bate2009,
       author = {{Bate}, Matthew R.},
        title = "{The importance of radiative feedback for the stellar initial mass function}",
      journal = {\mnras},
         year = 2009,
        month = feb,
       volume = {392},
       number = {4},
        pages = {1363-1380},
          doi = {10.1111/j.1365-2966.2008.14165.x},
archivePrefix = {arXiv},
       eprint = {0811.1035},
 primaryClass = {astro-ph},
       adsurl = {https://ui.adsabs.harvard.edu/abs/2009MNRAS.392.1363B}
}

@ARTICLE{2002ApJ...576..870P,
       author = {{Padoan}, Paolo and {Nordlund}, {\r{A}}ke},
        title = "{The Stellar Initial Mass Function from Turbulent Fragmentation}",
      journal = {\apj},
         year = 2002,
        month = sep,
       volume = {576},
       number = {2},
        pages = {870-879},
          doi = {10.1086/341790},
archivePrefix = {arXiv},
       eprint = {astro-ph/0011465},
 primaryClass = {astro-ph},
       adsurl = {https://ui.adsabs.harvard.edu/abs/2002ApJ...576..870P}
}

@ARTICLE{2018PASA...35...39H,
       author = {{Hopkins}, A.~M.},
        title = "{The Dawes Review 8: Measuring the Stellar Initial Mass Function}",
      journal = {\pasa},
         year = 2018,
        month = nov,
       volume = {35},
        pages = {39},
          doi = {10.1017/pasa.2018.29},
archivePrefix = {arXiv},
       eprint = {1807.09949},
 primaryClass = {astro-ph.GA},
       adsurl = {https://ui.adsabs.harvard.edu/abs/2018PASA...35...39H}
}

@INPROCEEDINGS{2014prpl.conf...53O,
       author = {{Offner}, S.~S.~R. and {Clark}, P.~C. and {Hennebelle}, P. and
         {Bastian}, N. and {Bate}, M.~R. and {Hopkins}, P.~F. and {Moraux}, E. and
         {Whitworth}, A.~P.},
        title = "{The Origin and Universality of the Stellar Initial Mass Function}",
    booktitle = {Protostars and Planets VI},
         year = 2014,
       editor = {{Beuther}, Henrik and {Klessen}, Ralf S. and {Dullemond}, Cornelis P. and
         {Henning}, Thomas},
        month = jan,
        pages = {53},
          doi = {10.2458/azu_uapress_9780816531240-ch003},
archivePrefix = {arXiv},
       eprint = {1312.5326},
 primaryClass = {astro-ph.SR},
       adsurl = {https://ui.adsabs.harvard.edu/abs/2014prpl.conf...53O}
}

@article{Haugb_lle_2018,
	doi = {10.3847/1538-4357/aaa432},
	url = {https://doi.org/10.3847%2F1538-4357%2Faaa432},
	year = 2018,
	month = {feb},
	publisher = {American Astronomical Society},
	volume = {854},
	number = {1},
	pages = {35},
	author = {Troels Haugb{\o}lle and Paolo Padoan and {\AA}ke Nordlund},
	title = {The Stellar {IMF} from Isothermal {MHD} Turbulence},
	journal = {The Astrophysical Journal}
}

@ARTICLE{2016MNRAS.462.4171B,
       author = {{Bertelli Motta}, C. and {Clark}, P.~C. and {Glover}, S.~C.~O. and
         {Klessen}, R.~S. and {Pasquali}, A.},
        title = "{The IMF as a function of supersonic turbulence}",
      journal = {\mnras},
         year = 2016,
        month = nov,
       volume = {462},
       number = {4},
        pages = {4171-4182},
          doi = {10.1093/mnras/stw1921},
archivePrefix = {arXiv},
       eprint = {1608.01306},
 primaryClass = {astro-ph.GA},
       adsurl = {https://ui.adsabs.harvard.edu/abs/2016MNRAS.462.4171B}
}

@ARTICLE{2017MNRAS.465..105L,
       author = {{Liptai}, David and {Price}, Daniel J. and {Wurster}, James and
         {Bate}, Matthew R.},
        title = "{Does turbulence determine the initial mass function?}",
      journal = {\mnras},
         year = 2017,
        month = feb,
       volume = {465},
       number = {1},
        pages = {105-110},
          doi = {10.1093/mnras/stw2770},
archivePrefix = {arXiv},
       eprint = {1610.07619},
 primaryClass = {astro-ph.GA},
       adsurl = {https://ui.adsabs.harvard.edu/abs/2017MNRAS.465..105L}
}

@ARTICLE{2016MNRAS.458..673G,
       author = {{Guszejnov}, D{\'a}vid and {Krumholz}, Mark R. and {Hopkins}, Philip F.},
        title = "{The necessity of feedback physics in setting the peak of the initial mass function}",
      journal = {\mnras},
         year = 2016,
        month = may,
       volume = {458},
       number = {1},
        pages = {673-680},
          doi = {10.1093/mnras/stw315},
archivePrefix = {arXiv},
       eprint = {1510.05040},
 primaryClass = {astro-ph.SR},
       adsurl = {https://ui.adsabs.harvard.edu/abs/2016MNRAS.458..673G}
}

@ARTICLE{2018A&A...611A..88L,
       author = {{Lee}, Yueh-Ning and {Hennebelle}, Patrick},
        title = "{Stellar mass spectrum within massive collapsing clumps. I. Influence of the initial conditions}",
      journal = {\aap},
         year = 2018,
        month = apr,
       volume = {611},
          eid = {A88},
        pages = {A88},
          doi = {10.1051/0004-6361/201731522},
archivePrefix = {arXiv},
       eprint = {1711.00316},
 primaryClass = {astro-ph.GA},
       adsurl = {https://ui.adsabs.harvard.edu/abs/2018A&A...611A..88L}
}

@ARTICLE{2019A&A...622A.125L,
       author = {{Lee}, Yueh-Ning and {Hennebelle}, Patrick},
        title = "{Stellar mass spectrum within massive collapsing clumps. III. Effects of temperature and magnetic field}",
      journal = {\aap},
         year = 2019,
        month = feb,
       volume = {622},
          eid = {A125},
        pages = {A125},
          doi = {10.1051/0004-6361/201834428},
archivePrefix = {arXiv},
       eprint = {1812.05508},
 primaryClass = {astro-ph.GA},
       adsurl = {https://ui.adsabs.harvard.edu/abs/2019A&A...622A.125L}
}

@INBOOK{2013pss5.book..115K,
       author = {{Kroupa}, Pavel and {Weidner}, Carsten and {Pflamm-Altenburg}, Jan and
         {Thies}, Ingo and {Dabringhausen}, J{\"o}rg and {Marks}, Michael and
         {Maschberger}, Thomas},
        title = "{The Stellar and Sub-Stellar Initial Mass Function of Simple and Composite Populations}",
    publisher = {Planets, Stars and Stellar Systems. Volume 5: Galactic Structure and Stellar Populations},
    booktitle = {Planets, Stars and Stellar Systems. Volume 5: Galactic Structure and Stellar Populations},
         year = 2013,
       volume = {5},
        pages = {115},
          doi = {10.1007/978-94-007-5612-0_4},
       adsurl = {https://ui.adsabs.harvard.edu/abs/2013pss5.book..115K}
}

@ARTICLE{2011JCoPh.230.3331W,
       author = {{Waagan}, K. and {Federrath}, C. and {Klingenberg}, C.},
        title = "{A robust numerical scheme for highly compressible magnetohydrodynamics: Nonlinear stability, implementation and tests}",
      journal = {Journal of Computational Physics},
         year = 2011,
        month = may,
       volume = {230},
       number = {9},
        pages = {3331-3351},
          doi = {10.1016/j.jcp.2011.01.026},
archivePrefix = {arXiv},
       eprint = {1101.3007},
 primaryClass = {astro-ph.IM},
       adsurl = {https://ui.adsabs.harvard.edu/abs/2011JCoPh.230.3331W}
}

@ARTICLE{mathew2020,
       author = {{Mathew}, Sajay Sunny and {Federrath}, Christoph},
        title = "{Implementation of stellar heating feedback in simulations of star cluster formation: effects on the initial mass function}",
      journal = {\mnras},
         year = 2020,
        month = jul,
       volume = {496},
       number = {4},
        pages = {5201-5210},
          doi = {10.1093/mnras/staa1931},
archivePrefix = {arXiv},
       eprint = {2007.01875},
 primaryClass = {astro-ph.GA},
       adsurl = {https://ui.adsabs.harvard.edu/abs/2020MNRAS.496.5201M}
}

@ARTICLE{hennebelle2020,
       author = {{Hennebelle}, Patrick and {Commer{\c{c}}on}, Beno{\^\i}t and {Lee}, Yueh-Ning and {Chabrier}, Gilles},
        title = "{What Is the Role of Stellar Radiative Feedback in Setting the Stellar Mass Spectrum?}",
      journal = {\apj},
         year = 2020,
        month = dec,
       volume = {904},
       number = {2},
          eid = {194},
        pages = {194},
          doi = {10.3847/1538-4357/abbfab},
archivePrefix = {arXiv},
       eprint = {2010.03539},
 primaryClass = {astro-ph.GA},
       adsurl = {https://ui.adsabs.harvard.edu/abs/2020ApJ...904..194H}
}

@ARTICLE{2009MNRAS.392..590B,
       author = {{Bate}, Matthew R.},
        title = "{Stellar, brown dwarf and multiple star properties from hydrodynamical simulations of star cluster formation}",
      journal = {\mnras},
         year = 2009,
        month = jan,
       volume = {392},
       number = {2},
        pages = {590-616},
          doi = {10.1111/j.1365-2966.2008.14106.x},
archivePrefix = {arXiv},
       eprint = {0811.0163},
 primaryClass = {astro-ph},
       adsurl = {https://ui.adsabs.harvard.edu/abs/2009MNRAS.392..590B}
}

@ARTICLE{2012ApJ...754...71K,
       author = {{Krumholz}, Mark R. and {Klein}, Richard I. and {McKee}, Christopher F.},
        title = "{Radiation-hydrodynamic Simulations of the Formation of Orion-like Star Clusters. II. The Initial Mass Function from Winds, Turbulence, and Radiation}",
      journal = {\apj},
         year = 2012,
        month = jul,
       volume = {754},
       number = {1},
          eid = {71},
        pages = {71},
          doi = {10.1088/0004-637X/754/1/71},
archivePrefix = {arXiv},
       eprint = {1203.2620},
 primaryClass = {astro-ph.SR},
       adsurl = {https://ui.adsabs.harvard.edu/abs/2012ApJ...754...71K}
}

@ARTICLE{2020MNRAS.497..336S,
       author = {{Sharda}, Piyush and {Federrath}, Christoph and {Krumholz}, Mark R.},
        title = "{The importance of magnetic fields for the initial mass function of the first stars}",
      journal = {\mnras},
         year = 2020,
        month = jul,
       volume = {497},
       number = {1},
        pages = {336-351},
          doi = {10.1093/mnras/staa1926},
archivePrefix = {arXiv},
       eprint = {2002.11502},
 primaryClass = {astro-ph.GA},
       adsurl = {https://ui.adsabs.harvard.edu/abs/2020MNRAS.497..336S}
}

@ARTICLE{1991A&A...248..485D,
       author = {{Duquennoy}, A. and {Mayor}, M.},
        title = "{Multiplicity among solar-type stars in the solar neighbourhood. II - Distribution of the orbital elements in an unbiased sample.}",
      journal = {\aap},
         year = 1991,
        month = aug,
       volume = {500},
        pages = {337-376},
       adsurl = {https://ui.adsabs.harvard.edu/abs/1991A&A...248..485D}
}

@INPROCEEDINGS{2004ASPC..318..166D,
       author = {{Delfosse}, X. and {Beuzit}, J. -L. and {Marchal}, L. and {Bonfils}, X. and
         {Perrier}, C. and {S{\'e}gransan}, D. and {Udry}, S. and {Mayor}, M. and
         {Forveille}, T.},
        title = "{M dwarfs binaries: Results from accurate radial velocities and high angular resolution observations}",
    booktitle = {Spectroscopically and Spatially Resolving the Components of the Close Binary Stars},
         year = 2004,
       editor = {{Hilditch}, R.~W. and {Hensberge}, H. and {Pavlovski}, K.},
       series = {Astronomical Society of the Pacific Conference Series},
       volume = {318},
        month = dec,
        pages = {166-174},
       adsurl = {https://ui.adsabs.harvard.edu/abs/2004ASPC..318..166D}
}

@ARTICLE{2014ApJ...788...40T,
       author = {{Todorov}, K.~O. and {Luhman}, K.~L. and {Konopacky}, Q.~M. and
         {McLeod}, K.~K. and {Apai}, D. and {Ghez}, A.~M. and {Pascucci}, I. and
         {Robberto}, M.},
        title = "{A Search for Companions to Brown Dwarfs in the Taurus and Chamaeleon Star-Forming Regions}",
      journal = {\apj},
         year = 2014,
        month = jun,
       volume = {788},
       number = {1},
          eid = {40},
        pages = {40},
          doi = {10.1088/0004-637X/788/1/40},
archivePrefix = {arXiv},
       eprint = {1404.0213},
 primaryClass = {astro-ph.GA},
       adsurl = {https://ui.adsabs.harvard.edu/abs/2014ApJ...788...40T}
}

@ARTICLE{2006AJ....132..663B,
       author = {{Basri}, Gibor and {Reiners}, Ansgar},
        title = "{A Survey for Spectroscopic Binaries among Very Low Mass Stars}",
      journal = {\aj},
         year = 2006,
        month = aug,
       volume = {132},
       number = {2},
        pages = {663-675},
          doi = {10.1086/505198},
archivePrefix = {arXiv},
       eprint = {astro-ph/0604259},
 primaryClass = {astro-ph},
       adsurl = {https://ui.adsabs.harvard.edu/abs/2006AJ....132..663B}
}

@ARTICLE{2003ApJ...587..407C,
       author = {{Close}, Laird M. and {Siegler}, Nick and {Freed}, Melanie and
         {Biller}, Beth},
        title = "{Detection of Nine M8.0-L0.5 Binaries: The Very Low Mass Binary Population and Its Implications for Brown Dwarf and Very Low Mass Star Formation}",
      journal = {\apj},
         year = 2003,
        month = apr,
       volume = {587},
       number = {1},
        pages = {407-422},
          doi = {10.1086/368177},
archivePrefix = {arXiv},
       eprint = {astro-ph/0301095},
 primaryClass = {astro-ph},
       adsurl = {https://ui.adsabs.harvard.edu/abs/2003ApJ...587..407C}
}

@ARTICLE{1992ApJ...396..178F,
       author = {{Fischer}, Debra A. and {Marcy}, Geoffrey W.},
        title = "{Multiplicity among M Dwarfs}",
      journal = {\apj},
         year = 1992,
        month = sep,
       volume = {396},
        pages = {178},
          doi = {10.1086/171708},
       adsurl = {https://ui.adsabs.harvard.edu/abs/1992ApJ...396..178F}
}

@ARTICLE{2010ApJS..190....1R,
       author = {{Raghavan}, Deepak and {McAlister}, Harold A. and {Henry}, Todd J. and
         {Latham}, David W. and {Marcy}, Geoffrey W. and {Mason}, Brian D. and
         {Gies}, Douglas R. and {White}, Russel J. and {ten Brummelaar}, Theo A.},
        title = "{A Survey of Stellar Families: Multiplicity of Solar-type Stars}",
      journal = {\apjs},
         year = 2010,
        month = sep,
       volume = {190},
       number = {1},
        pages = {1-42},
          doi = {10.1088/0067-0049/190/1/1},
archivePrefix = {arXiv},
       eprint = {1007.0414},
 primaryClass = {astro-ph.SR},
       adsurl = {https://ui.adsabs.harvard.edu/abs/2010ApJS..190....1R}
}

@ARTICLE{1986ApJ...309..275H,
       author = {{Hartmann}, L. and {Hewett}, R. and {Stahler}, S. and {Mathieu}, R.~D.},
        title = "{Rotational and Radial Velocities of T Tauri Stars}",
      journal = {\apj},
         year = 1986,
        month = oct,
       volume = {309},
        pages = {275},
          doi = {10.1086/164599},
       adsurl = {https://ui.adsabs.harvard.edu/abs/1986ApJ...309..275H}
}

@ARTICLE{2000ApJ...543..822B,
       author = {{Burkert}, Andreas and {Bodenheimer}, Peter},
        title = "{Turbulent Molecular Cloud Cores: Rotational Properties}",
      journal = {\apj},
         year = 2000,
        month = nov,
       volume = {543},
       number = {2},
        pages = {822-830},
          doi = {10.1086/317122},
archivePrefix = {arXiv},
       eprint = {astro-ph/0006010},
 primaryClass = {astro-ph},
       adsurl = {https://ui.adsabs.harvard.edu/abs/2000ApJ...543..822B}
}

@INPROCEEDINGS{1992ASPC...32...41S,
       author = {{Simon}, M.},
        title = "{Multiplicity Among the Young Stars}",
    booktitle = {IAU Colloq. 135: Complementary Approaches to Double and Multiple Star Research},
         year = 1992,
       editor = {{McAlister}, H.~A. and {Hartkopf}, W.~I.},
       series = {Astronomical Society of the Pacific Conference Series},
       volume = {32},
        month = jan,
        pages = {41},
       adsurl = {https://ui.adsabs.harvard.edu/abs/1992ASPC...32...41S}
}

@ARTICLE{1993ApJ...406..528G,
       author = {{Goodman}, A.~A. and {Benson}, P.~J. and {Fuller}, G.~A. and
         {Myers}, P.~C.},
        title = "{Dense Cores in Dark Clouds. VIII. Velocity Gradients}",
      journal = {\apj},
         year = 1993,
        month = apr,
       volume = {406},
        pages = {528},
          doi = {10.1086/172465},
       adsurl = {https://ui.adsabs.harvard.edu/abs/1993ApJ...406..528G}
}

@ARTICLE{2015ApJ...812..129Y,
       author = {{Yen}, Hsi-Wei and {Takakuwa}, Shigehisa and {Koch}, Patrick M. and
         {Aso}, Yusuke and {Koyamatsu}, Shin and {Krasnopolsky}, Ruben and
         {Ohashi}, Nagayoshi},
        title = "{No Keplerian Disk >10 AU Around the Protostar B335: Magnetic Braking or Young Age?}",
      journal = {\apj},
         year = 2015,
        month = oct,
       volume = {812},
       number = {2},
          eid = {129},
        pages = {129},
          doi = {10.1088/0004-637X/812/2/129},
archivePrefix = {arXiv},
       eprint = {1509.04675},
 primaryClass = {astro-ph.SR},
       adsurl = {https://ui.adsabs.harvard.edu/abs/2015ApJ...812..129Y}
}

@ARTICLE{2004A&A...423....1J,
       author = {{Jappsen}, A. -K. and {Klessen}, R.~S.},
        title = "{Protostellar angular momentum evolution  during gravoturbulent fragmentation}",
      journal = {\aap},
         year = 2004,
        month = aug,
       volume = {423},
        pages = {1-12},
          doi = {10.1051/0004-6361:20040220},
archivePrefix = {arXiv},
       eprint = {astro-ph/0402361},
 primaryClass = {astro-ph},
       adsurl = {https://ui.adsabs.harvard.edu/abs/2004A&A...423....1J}
}

@ARTICLE{2010ApJ...709...27W,
       author = {{Wang}, Peng and {Li}, Zhi-Yun and {Abel}, Tom and {Nakamura}, Fumitaka},
        title = "{Outflow Feedback Regulated Massive Star Formation in Parsec-Scale Cluster-Forming Clumps}",
      journal = {\apj},
         year = 2010,
        month = jan,
       volume = {709},
       number = {1},
        pages = {27-41},
          doi = {10.1088/0004-637X/709/1/27},
archivePrefix = {arXiv},
       eprint = {0908.4129},
 primaryClass = {astro-ph.SR},
       adsurl = {https://ui.adsabs.harvard.edu/abs/2010ApJ...709...27W}
}

@ARTICLE{2008ApJ...684..395H,
       author = {{Hennebelle}, Patrick and {Chabrier}, Gilles},
        title = "{Analytical Theory for the Initial Mass Function: CO Clumps and Prestellar Cores}",
      journal = {\apj},
         year = 2008,
        month = sep,
       volume = {684},
       number = {1},
        pages = {395-410},
          doi = {10.1086/589916},
archivePrefix = {arXiv},
       eprint = {0805.0691},
 primaryClass = {astro-ph},
       adsurl = {https://ui.adsabs.harvard.edu/abs/2008ApJ...684..395H}
}

@ARTICLE{2009ApJ...702.1428H,
       author = {{Hennebelle}, Patrick and {Chabrier}, Gilles},
        title = "{Analytical Theory for the Initial Mass Function. II. Properties of the Flow}",
      journal = {\apj},
         year = 2009,
        month = sep,
       volume = {702},
       number = {2},
        pages = {1428-1442},
          doi = {10.1088/0004-637X/702/2/1428},
archivePrefix = {arXiv},
       eprint = {0907.2765},
 primaryClass = {astro-ph.GA},
       adsurl = {https://ui.adsabs.harvard.edu/abs/2009ApJ...702.1428H}
}

@ARTICLE{2012MNRAS.423.2037H,
       author = {{Hopkins}, Philip F.},
        title = "{The stellar initial mass function, core mass function and the last-crossing distribution}",
      journal = {\mnras},
         year = 2012,
        month = jul,
       volume = {423},
       number = {3},
        pages = {2037-2044},
          doi = {10.1111/j.1365-2966.2012.20731.x},
archivePrefix = {arXiv},
       eprint = {1201.4387},
 primaryClass = {astro-ph.GA},
       adsurl = {https://ui.adsabs.harvard.edu/abs/2012MNRAS.423.2037H}
}

@ARTICLE{2013MNRAS.430.1653H,
       author = {{Hopkins}, Philip F.},
        title = "{A general theory of turbulent fragmentation}",
      journal = {\mnras},
         year = 2013,
        month = apr,
       volume = {430},
       number = {3},
        pages = {1653-1693},
          doi = {10.1093/mnras/sts704},
archivePrefix = {arXiv},
       eprint = {1210.0903},
 primaryClass = {astro-ph.CO},
       adsurl = {https://ui.adsabs.harvard.edu/abs/2013MNRAS.430.1653H}
}

@ARTICLE{2022MNRAS.512..216G,
       author = {{Grudi{\'c}}, Michael Y. and {Guszejnov}, D{\'a}vid and {Offner}, Stella S.~R. and {Rosen}, Anna L. and {Raju}, Aman N. and {Faucher-Gigu{\`e}re}, Claude-Andr{\'e} and {Hopkins}, Philip F.},
        title = "{The dynamics and outcome of star formation with jets, radiation, winds, and supernovae in concert}",
      journal = {\mnras},
         year = 2022,
        month = may,
       volume = {512},
       number = {1},
        pages = {216-232},
          doi = {10.1093/mnras/stac526},
archivePrefix = {arXiv},
       eprint = {2201.00882},
 primaryClass = {astro-ph.GA},
       adsurl = {https://ui.adsabs.harvard.edu/abs/2022MNRAS.512..216G}
}

@ARTICLE{2018MNRAS.473.4975T,
       author = {{Traficante}, A. and {Fuller}, G.~A. and {Smith}, R.~J. and {Billot}, N. and {Duarte-Cabral}, A. and {Peretto}, N. and {Molinari}, S. and {Pineda}, J.~E.},
        title = "{Massive 70 {\ensuremath{\mu}}m quiet clumps - II. Non-thermal motions driven by gravity in massive star formation?}",
      journal = {\mnras},
         year = 2018,
        month = feb,
       volume = {473},
       number = {4},
        pages = {4975-4985},
          doi = {10.1093/mnras/stx2672},
archivePrefix = {arXiv},
       eprint = {1710.04904},
 primaryClass = {astro-ph.GA},
       adsurl = {https://ui.adsabs.harvard.edu/abs/2018MNRAS.473.4975T}
}

@ARTICLE{2025MNRAS.tmp.1946V,
       author = {{V{\'a}zquez-Semadeni}, Enrique and {Palau}, Aina and {G{\'o}mez}, Gilberto C. and {Arroyo-Ch{\'a}vez}, Griselda and {Alig}, Christian and {Ballesteros-Paredes}, Javier and {Camacho}, Vianey and {Traficante}, Alessio and {Gonz{\'a}lez-Samaniego}, Alejandro and {Zamora-Avil{\'e}s}, Manuel and {Burkert}, Andreas},
        title = "{The Turbulent Support (TS) and Global Hierarchical Collapse (GHC) models for molecular clouds compared. Differences, convergence, and myths}",
      journal = {\mnras},
         year = 2025,
        month = nov,
          doi = {10.1093/mnras/staf2059},
archivePrefix = {arXiv},
       eprint = {2408.10406},
 primaryClass = {astro-ph.GA},
       adsurl = {https://ui.adsabs.harvard.edu/abs/2025MNRAS.tmp.1946V}
}

@ARTICLE{2026A&A...706A.178N,
       author = {{N{\'u}{\~n}ez-Casti{\~n}eyra}, A. and {Gonz{\'a}lez}, M. and {Brucy}, N. and {Hennebelle}, P. and {Louvet}, F. and {Motte}, F.},
        title = "{The interdependence between density PDF, CMF, and IMF and their relation with Mach number in simulations}",
      journal = {\aap},
         year = 2026,
        month = feb,
       volume = {706},
          eid = {A178},
        pages = {A178},
          doi = {10.1051/0004-6361/202453497},
archivePrefix = {arXiv},
       eprint = {2412.12809},
 primaryClass = {astro-ph.GA},
       adsurl = {https://ui.adsabs.harvard.edu/abs/2026A&A...706A.178N}
}

@ARTICLE{2021MNRAS.506.3239G,
       author = {{Grudi{\'c}}, Michael Y. and {Kruijssen}, J.~M. Diederik and {Faucher-Gigu{\`e}re}, Claude-Andr{\'e} and {Hopkins}, Philip F. and {Ma}, Xiangcheng and {Quataert}, Eliot and {Boylan-Kolchin}, Michael},
        title = "{A model for the formation of stellar associations and clusters from giant molecular clouds}",
      journal = {\mnras},
         year = 2021,
        month = sep,
       volume = {506},
       number = {3},
        pages = {3239-3258},
          doi = {10.1093/mnras/stab1894},
archivePrefix = {arXiv},
       eprint = {2008.04453},
 primaryClass = {astro-ph.GA},
       adsurl = {https://ui.adsabs.harvard.edu/abs/2021MNRAS.506.3239G}
}

@ARTICLE{2016MNRAS.458.1671K,
       author = {{Krumholz}, Mark R. and {Burkhart}, Blakesley},
        title = "{Is turbulence in the interstellar medium driven by feedback or gravity? An observational test}",
      journal = {\mnras},
         year = 2016,
        month = may,
       volume = {458},
       number = {2},
        pages = {1671-1677},
          doi = {10.1093/mnras/stw434},
archivePrefix = {arXiv},
       eprint = {1512.03439},
 primaryClass = {astro-ph.GA},
       adsurl = {https://ui.adsabs.harvard.edu/abs/2016MNRAS.458.1671K}
}

@ARTICLE{2020MNRAS.493.2872C,
       author = {{Chevance}, M{\'e}lanie and {Kruijssen}, J.~M. Diederik and {Hygate}, Alexander P.~S. and {Schruba}, Andreas and {Longmore}, Steven N. and {Groves}, Brent and {Henshaw}, Jonathan D. and {Herrera}, Cinthya N. and {Hughes}, Annie and {Jeffreson}, Sarah M.~R. and {Lang}, Philipp and {Leroy}, Adam K. and {Meidt}, Sharon E. and {Pety}, J{\'e}r{\^o}me and {Razza}, Alessandro and {Rosolowsky}, Erik and {Schinnerer}, Eva and {Bigiel}, Frank and {Blanc}, Guillermo A. and {Emsellem}, Eric and {Faesi}, Christopher M. and {Glover}, Simon C.~O. and {Haydon}, Daniel T. and {Ho}, I.-Ting and {Kreckel}, Kathryn and {Lee}, Janice C. and {Liu}, Daizhong and {Querejeta}, Miguel and {Saito}, Toshiki and {Sun}, Jiayi and {Usero}, Antonio and {Utomo}, Dyas},
        title = "{The lifecycle of molecular clouds in nearby star-forming disc galaxies}",
      journal = {\mnras},
         year = 2020,
        month = apr,
       volume = {493},
       number = {2},
        pages = {2872-2909},
          doi = {10.1093/mnras/stz3525},
archivePrefix = {arXiv},
       eprint = {1911.03479},
 primaryClass = {astro-ph.GA},
       adsurl = {https://ui.adsabs.harvard.edu/abs/2020MNRAS.493.2872C}
}

@ARTICLE{2009ApJS..181..321E,
       author = {{Evans}, Neal J., II and {Dunham}, Michael M. and {J{\o}rgensen}, Jes K. and {Enoch}, Melissa L. and {Mer{\'\i}n}, Bruno and {van Dishoeck}, Ewine F. and {Alcal{\'a}}, Juan M. and {Myers}, Philip C. and {Stapelfeldt}, Karl R. and {Huard}, Tracy L. and {Allen}, Lori E. and {Harvey}, Paul M. and {van Kempen}, Tim and {Blake}, Geoffrey A. and {Koerner}, David W. and {Mundy}, Lee G. and {Padgett}, Deborah L. and {Sargent}, Anneila I.},
        title = "{The Spitzer c2d Legacy Results: Star-Formation Rates and Efficiencies; Evolution and Lifetimes}",
      journal = {\apjs},
         year = 2009,
        month = apr,
       volume = {181},
       number = {2},
        pages = {321-350},
          doi = {10.1088/0067-0049/181/2/321},
archivePrefix = {arXiv},
       eprint = {0811.1059},
 primaryClass = {astro-ph},
       adsurl = {https://ui.adsabs.harvard.edu/abs/2009ApJS..181..321E}
}

@ARTICLE{sfr_km2005,
       author = {{Krumholz}, Mark R. and {McKee}, Christopher F.},
        title = "{A General Theory of Turbulence-regulated Star Formation, from Spirals to Ultraluminous Infrared Galaxies}",
      journal = {\apj},
         year = 2005,
        month = sep,
       volume = {630},
       number = {1},
        pages = {250-268},
          doi = {10.1086/431734},
archivePrefix = {arXiv},
       eprint = {astro-ph/0505177},
 primaryClass = {astro-ph},
       adsurl = {https://ui.adsabs.harvard.edu/abs/2005ApJ...630..250K}
}

@ARTICLE{2016ApJ...833..229L,
       author = {{Lee}, Eve J. and {Miville-Desch{\^e}nes}, Marc-Antoine and {Murray}, Norman W.},
        title = "{Observational Evidence of Dynamic Star Formation Rate in Milky Way Giant Molecular Clouds}",
      journal = {\apj},
         year = 2016,
        month = dec,
       volume = {833},
       number = {2},
          eid = {229},
        pages = {229},
          doi = {10.3847/1538-4357/833/2/229},
archivePrefix = {arXiv},
       eprint = {1608.05415},
 primaryClass = {astro-ph.GA},
       adsurl = {https://ui.adsabs.harvard.edu/abs/2016ApJ...833..229L}
}

@ARTICLE{2011ApJ...729..133M,
       author = {{Murray}, Norman},
        title = "{Star Formation Efficiencies and Lifetimes of Giant Molecular Clouds in the Milky Way}",
      journal = {\apj},
         year = 2011,
        month = mar,
       volume = {729},
       number = {2},
          eid = {133},
        pages = {133},
          doi = {10.1088/0004-637X/729/2/133},
archivePrefix = {arXiv},
       eprint = {1007.3270},
 primaryClass = {astro-ph.GA},
       adsurl = {https://ui.adsabs.harvard.edu/abs/2011ApJ...729..133M}
}

@ARTICLE{sfr_hc2011,
       author = {{Hennebelle}, Patrick and {Chabrier}, Gilles},
        title = "{Analytical Star Formation Rate from Gravoturbulent Fragmentation}",
      journal = {\apjl},
         year = 2011,
        month = dec,
       volume = {743},
       number = {2},
          eid = {L29},
        pages = {L29},
          doi = {10.1088/2041-8205/743/2/L29},
archivePrefix = {arXiv},
       eprint = {1110.0033},
 primaryClass = {astro-ph.GA},
       adsurl = {https://ui.adsabs.harvard.edu/abs/2011ApJ...743L..29H}
}

@ARTICLE{2016ApJ...831...73V,
       author = {{Vutisalchavakul}, Nalin and {Evans}, Neal J., II and {Heyer}, Mark},
        title = "{Star Formation Relations in the Milky Way}",
      journal = {\apj},
         year = 2016,
        month = nov,
       volume = {831},
       number = {1},
          eid = {73},
        pages = {73},
          doi = {10.3847/0004-637X/831/1/73},
archivePrefix = {arXiv},
       eprint = {1607.06518},
 primaryClass = {astro-ph.SR},
       adsurl = {https://ui.adsabs.harvard.edu/abs/2016ApJ...831...73V}
}

@ARTICLE{2020A&A...637A..92G,
       author = {{Gaudel}, M. and {Maury}, A.~J. and {Belloche}, A. and {Maret}, S. and {Andr{\'e}}, Ph. and {Hennebelle}, P. and {Galametz}, M. and {Testi}, L. and {Cabrit}, S. and {Palmeirim}, P. and {Ladjelate}, B. and {Codella}, C. and {Podio}, L.},
        title = "{Angular momentum profiles of Class 0 protostellar envelopes}",
      journal = {\aap},
         year = 2020,
        month = may,
       volume = {637},
          eid = {A92},
        pages = {A92},
          doi = {10.1051/0004-6361/201936364},
archivePrefix = {arXiv},
       eprint = {2001.10004},
 primaryClass = {astro-ph.SR},
       adsurl = {https://ui.adsabs.harvard.edu/abs/2020A&A...637A..92G}
}

@ARTICLE{2015ApJ...799..193Y,
       author = {{Yen}, Hsi-Wei and {Koch}, Patrick M. and {Takakuwa}, Shigehisa and {Ho}, Paul T.~P. and {Ohashi}, Nagayoshi and {Tang}, Ya-Wen},
        title = "{Observations of Infalling and Rotational Motions on a 1000 AU Scale around 17 Class 0 and 0/I Protostars: Hints of Disk Growth and Magnetic Braking?}",
      journal = {\apj},
         year = 2015,
        month = feb,
       volume = {799},
       number = {2},
          eid = {193},
        pages = {193},
          doi = {10.1088/0004-637X/799/2/193},
archivePrefix = {arXiv},
       eprint = {1412.1916},
 primaryClass = {astro-ph.SR},
       adsurl = {https://ui.adsabs.harvard.edu/abs/2015ApJ...799..193Y}
}

@ARTICLE{1997ApJ...488..317O,
       author = {{Ohashi}, Nagayoshi and {Hayashi}, Masahiko and {Ho}, Paul. T.~P. and {Momose}, Munetake and {Tamura}, Motohide and {Hirano}, Naomi and {Sargent}, Anneila I.},
        title = "{Rotation in the Protostellar Envelopes around IRAS 04169+2702 and IRAS 04365+2535: The Size Scale for Dynamical Collapse}",
      journal = {\apj},
         year = 1997,
        month = oct,
       volume = {488},
       number = {1},
        pages = {317-329},
          doi = {10.1086/304685},
       adsurl = {https://ui.adsabs.harvard.edu/abs/1997ApJ...488..317O}
}

@ARTICLE{2002ApJ...572..238C,
       author = {{Caselli}, Paola and {Benson}, Priscilla J. and {Myers}, Philip C. and {Tafalla}, Mario},
        title = "{Dense Cores in Dark Clouds. XIV. N$_{2}$H$^{+}$ (1-0) Maps of Dense Cloud Cores}",
      journal = {\apj},
         year = 2002,
        month = jun,
       volume = {572},
       number = {1},
        pages = {238-263},
          doi = {10.1086/340195},
archivePrefix = {arXiv},
       eprint = {astro-ph/0202173},
 primaryClass = {astro-ph},
       adsurl = {https://ui.adsabs.harvard.edu/abs/2002ApJ...572..238C}
}

@ARTICLE{1994ApJ...423..681V,
       author = {{Vazquez-Semadeni}, Enrique},
        title = "{Hierarchical Structure in Nearly Pressureless Flows as a Consequence of Self-similar Statistics}",
      journal = {\apj},
         year = 1994,
        month = mar,
       volume = {423},
        pages = {681},
          doi = {10.1086/173847},
       adsurl = {https://ui.adsabs.harvard.edu/abs/1994ApJ...423..681V}
}

@ARTICLE{2007ApJ...665..416K,
       author = {{Kritsuk}, Alexei G. and {Norman}, Michael L. and {Padoan}, Paolo and {Wagner}, Rick},
        title = "{The Statistics of Supersonic Isothermal Turbulence}",
      journal = {\apj},
         year = 2007,
        month = aug,
       volume = {665},
       number = {1},
        pages = {416-431},
          doi = {10.1086/519443},
archivePrefix = {arXiv},
       eprint = {0704.3851},
 primaryClass = {astro-ph},
       adsurl = {https://ui.adsabs.harvard.edu/abs/2007ApJ...665..416K}
}

@ARTICLE{1997MNRAS.288..145P,
       author = {{Padoan}, Paolo and {Nordlund}, Ake and {Jones}, Bernard J.~T.},
        title = "{The universality of the stellar initial mass function}",
      journal = {\mnras},
         year = 1997,
        month = jun,
       volume = {288},
       number = {1},
        pages = {145-152},
          doi = {10.1093/mnras/288.1.145},
archivePrefix = {arXiv},
       eprint = {astro-ph/9703110},
 primaryClass = {astro-ph},
       adsurl = {https://ui.adsabs.harvard.edu/abs/1997MNRAS.288..145P}
}

@ARTICLE{1995MNRAS.277..377P,
       author = {{Padoan}, Paolo},
        title = "{Supersonic turbulent flows and the fragmentation of a cold medium}",
      journal = {\mnras},
         year = 1995,
        month = nov,
       volume = {277},
       number = {2},
        pages = {377-388},
          doi = {10.1093/mnras/277.2.377},
archivePrefix = {arXiv},
       eprint = {astro-ph/9506002},
 primaryClass = {astro-ph},
       adsurl = {https://ui.adsabs.harvard.edu/abs/1995MNRAS.277..377P}
}

@ARTICLE{1993ApJ...419L..29E,
       author = {{Elmegreen}, B.~G.},
        title = "{Star Formation at Compressed Interfaces in Turbulent Self-gravitating Clouds}",
      journal = {\apjl},
         year = 1993,
        month = dec,
       volume = {419},
        pages = {L29},
          doi = {10.1086/187129},
       adsurl = {https://ui.adsabs.harvard.edu/abs/1993ApJ...419L..29E}
}

@ARTICLE{2010A&A...516A..25S,
       author = {{Schmidt}, W. and {Kern}, S.~A.~W. and {Federrath}, C. and {Klessen}, R.~S.},
        title = "{Numerical and semi-analytic core mass distributions in supersonic isothermal turbulence}",
      journal = {\aap},
         year = 2010,
        month = jun,
       volume = {516},
          eid = {A25},
        pages = {A25},
          doi = {10.1051/0004-6361/200913904},
archivePrefix = {arXiv},
       eprint = {1002.2359},
 primaryClass = {astro-ph.SR},
       adsurl = {https://ui.adsabs.harvard.edu/abs/2010A&A...516A..25S}
}




\appendix
\section{Relationship between the dimensionless parameters}
\label{sec: isolate_proof}
The following equations show that $\mathcal{M}$, $\alpha_{\mathrm{vir}}$, and $M_{\mathrm{cl}}/M_{\mathrm{J}}$ are interconnected, making it difficult to fully isolate the individual effects of these quantities through numerical experiments. To show this, we start by writing down the main definitions of the dimensionless parameters.

\paragraph*{Mach number:}
\begin{equation}
    \mathcal{M} = \sigma_{\mathrm{v}}/c_{\mathrm{s}}.
    \label{eq: mach}
\end{equation}

\paragraph*{Number of Jeans masses:}
\begin{equation}
    \frac{M_{\mathrm{cl}}}{M_{\mathrm{J}}} = \frac{\rho_0L^3}{\frac{4}{3}\pi\rho_0(l_{\mathrm{J}}/2)^3} = \frac{6G^{3/2}{\rho_0}^{3/2}L^3}{\pi^{5/2}{c_{\mathrm{s}}^3}},
\end{equation}
where
the Jeans length at the mean density is
\begin{equation*}
    l_{\mathrm{J}} = \sqrt{\pi c_{\mathrm{s}}^2/(G \rho_0)}.
\end{equation*}
Note that we have defined the total cloud mass as $M_{\mathrm{cl}} = \rho_0 L^3$ since our cloud is represented by a cubic computational domain. Changing this to spherical geometry merely changes the coefficients, but does not change the dependencies.

\paragraph*{Virial parameter:}
\begin{equation}
    \alpha_{\mathrm{vir}} = \frac{5\sigma_\mathrm{v}^2L}{6GM_{\mathrm{cl}}} =
    \frac{5\mathcal{M}^2c_{\mathrm{s}}^2}{6G\rho_0 L^2} = \frac{5\mathcal{M}^2}{6^{1/3}\pi^{5/3}{(M_{\mathrm{cl}}/M_{\mathrm{J}})}^{2/3}}.
    \label{eq:avir_as_func_of_mach_and_nj}
\end{equation}
Rearranging Eq.~(\ref{eq:avir_as_func_of_mach_and_nj}) for $\mathcal{M}$ yields
\begin{equation}
    \mathcal{M} = \frac{6^{1/6}\pi^{5/6}\alpha_{\mathrm{vir}}^{1/2}(M_{\mathrm{cl}}/M_{\mathrm{J}})^{1/3}}{5^{1/2}},
\end{equation}
or rearranging for $M_{\mathrm{cl}}/M_{\mathrm{J}}$ yields
\begin{equation}
    \frac{M_{\mathrm{cl}}}{M_{\mathrm{J}}} = \frac{5^{3/2}\mathcal{M}^3}{6^{1/2}\pi^{5/2}\alpha_{\mathrm{vir}}^{3/2}}.
\end{equation}
Thus, it is clear that, when varying $\mathcal{M}$, the number of Jeans masses $M_{\mathrm{cl}}/M_{\mathrm{J}}$ also changes if one wants to keep $\alpha_{\mathrm{vir}}$ constant. This interdependence means that only two of these three parameters can be varied independently.


\bsp	
\label{lastpage}
\end{document}